\documentclass[10pt]{amsart}
\usepackage[margin=2.5cm]{geometry}

\usepackage{hyperref}
\usepackage{mathtools}  
\usepackage{mathrsfs,amsfonts,graphicx}
\usepackage{amsmath}
\usepackage{stackrel,amssymb}
\usepackage{textcomp}
\usepackage{array}
\usepackage[usenames, dvipsnames]{color}
\usepackage[font=small]{caption}
\usepackage{floatrow}
\usepackage{subfig}
\usepackage{cite}
\usepackage[utf8]{inputenc}
\usepackage{amsthm}
\usepackage{multicol}
\usepackage{tikz}
\usepackage{bbm}
\usetikzlibrary{calc,arrows}

\usepackage{multirow}

\numberwithin{equation}{section}

\newcommand{\be}{\begin{equation}}
\newcommand{\ee}{\end{equation}}
\newcommand{\nnbe}{\begin{equation*}}
\newcommand{\nnee}{\end{equation*}}
\newcommand{\bea}{\begin{eqnarray}}
\newcommand{\eea}{\end{eqnarray}}
\newcommand{\ba}{\begin{align}}
\newcommand{\ea}{\end{align}}
\newcommand{\bi}{\begin{itemize}}
\newcommand{\ei}{\end{itemize}}
\newtheorem{thm}{Theorem}[section]
\theoremstyle{definition}
\newtheorem{defn}[thm]{Definition}
\theoremstyle{plain}
\newtheorem{prp}[thm]{Proposition}
\newtheorem{crl}[thm]{Corollary}

\theoremstyle{definition}
\newtheorem{rmk}[thm]{Remark}
\theoremstyle{definition}
\newtheorem{ex}[thm]{Example}

\theoremstyle{plain}

\newsavebox{\overlongequation}

\newcommand{\eref}[1]{Equation~\eqref{#1}}
\newcommand{\erefs}[1]{Equations~\eqref{#1}}
\newcommand{\sref}[1]{Section~\ref{#1}}
\newcommand{\srefs}[1]{Sections~\ref{#1}}

\newcommand{\bnd}{\mathcal{B}}			
\newcommand{\ec}{\chi}					
\newcommand{\fl}[1]{{#1}'}					
\newcommand{\contr}[1]{{#1}_\mathrm{sing}}		
\newcommand{\defm}[1]{{#1}^\flat}				

\title[Flops for Complete Intersection Calabi-Yau Threefolds]{Flops for Complete Intersection Calabi-Yau Threefolds}

\author[Brodie]{Callum Brodie}
\address{Department of Physics, Robeson Hall, Virginia Tech, Blacksburg, VA 24061, USA}
\email{callumb@vt.edu}

\author[Constantin]{Andrei Constantin}
\address{Rudolf Peierls Centre for Theoretical Physics\\
University of Oxford\\
Parks Road\\
Oxford, OX1 3PU
\\UK}
\email{andrei.constantin@physics.ox.ac.uk}

\author[Lukas]{Andre Lukas}
\address{Rudolf Peierls Centre for Theoretical Physics\\
University of Oxford\\
Parks Road\\
Oxford, OX1 3PU
\\UK}
\email{andre.lukas@physics.ox.ac.uk}

\author[Ruehle]{Fabian Ruehle}
\address{Department of Physics \& Department of Mathematics, Northeastern University, 360 Huntington Avenue, Boston MA 02115, USA}
\address{The NSF AI Institute for Artificial Intelligence and Fundamental Interactions}
\email{f.ruehle@northeastern.edu}

\begin{document}

\begin{abstract}\noindent
We study flops of Calabi-Yau threefolds realised as K\"ahler-favourable complete intersections in products of projective spaces (CICYs) and identify two different types. The existence and the type of the flops can be recognised from the configuration matrix of the CICY, which also allows for constructing such examples. The first type corresponds to rows containing only $1$s and $0$s, while the second type corresponds to rows containing a single entry of $2$, followed by $1$s and $0$s. We give explicit descriptions for the manifolds obtained after the flop and show that the second type of flop always leads to isomorphic manifolds, while the first type in general leads to non-isomorphic flops. The singular manifolds involved in the flops are determinantal varieties in the first case and more complicated in the second case. We also discuss manifolds admitting an infinite chain of flops and show how to identify these from the configuration matrix. Finally, we point out how to construct the divisor images and Picard group isomorphisms under both types of flops.

\end{abstract}

\maketitle

\tableofcontents

\setcounter{footnote}{0}

\section{Introduction}
Flops play an important role in the classification of three-dimensional varieties, as well as in the study of topological transitions in string theory. Due to a theorem of Kawamata \cite{Kawamata:1988}, it is known that all birational maps between three-dimensional minimal models can be expressed as compositions of simple flops in which a single irreducible curve is changed. In string theory flops correspond to mild topological transitions that could in principle be dynamically realised in the very early stages of the cosmological evolution.  

Locally, a simple flop corresponds to a codimension two surgery, in which a $\mathbb P^1$ curve is collapsed and replaced by a $\mathbb P^1$ in a ``transverse'' direction. On a smooth threefold a general flop can be locally deformed into a disjoint union of simple flops. The best known example of a simple flop is due to Atiyah~\cite{Atiyah:1958}, which we review here. The discussion starts with a singular hypersurface $X_{\rm sing}\subset \mathbb C^4$ defined by the equation
\begin{equation*}
xy - zt = 0 ~.
\end{equation*}
The point $(x,y,z,t)=(0,0,0,0)$ is an ordinary double point (an $A_1$ singularity) and can be blown-up to $\mathbb P^1\times \mathbb P^1$. However, this is not the smallest possible resolution. Instead of blowing-up the singular point directly, one can blow-up along the line $y=t=0$, which replaces the origin $(0,0,0,0)$ with a $\mathbb P^1$, producing a non-singular threefold $X^+$ described by the following two equations in $\mathbb C^4\times \mathbb{P}^1$
\be
\left(\begin{array}{cc} x&z \\t&y\end{array}\right) \left(\begin{array}{c} u_1\\u_2\end{array}\right)  = \vec 0~,
\ee
where $[u_1:u_2]$ are homogeneous coordinates on $\mathbb{P}^1$. 
Another way of resolving $X_{\rm sing}$ is to blow-up along the line $x=z=0$, which produces another non-singular threefold $X^-$, described by two other equations  in $\mathbb C^4\times \mathbb{P}^1$,
\be
\left(\begin{array}{cc} x&t \\z&y\end{array}\right) \left(\begin{array}{c} v_1\\v_2\end{array}\right)  = \vec 0~.
\ee
where $[v_1:v_2]$ are homogeneous coordinates on $\mathbb{P}^1$. 
The threefold $X^-$ is isomorphic to $X^+$ away from the~$\mathbb{P}^1$ and hence isomorphic in codimension one. The Atiyah flop is the birational map relating $X^+$ and $X^-$. The example shows that there is no minimal resolution of $X_{\rm sing}$, but rather two small resolutions related by a flop. 

More generally, Koll\'ar and Mori have shown~\cite{Kollar:1989} that the singular manifold $X_{\rm sing}$ arising in a flop has a hypersurface singularity and hence can be locally described as $f=0$, where in appropriate coordinates $f$ vanishes at the origin and
\begin{equation}\label{eq:local_flop}
f (x)= x_1^2 + g(x_2,x_3,x_4)~.
\end{equation}
The flop is then induced by the automorphism $x_1 \rightarrow -x_1$ of the singular threefold $X_{\rm sing}$. The Atiyah flop corresponds to the function $f (x)= x_1^2 + x_2^2+x_3^2+x_4^2$, while a flop involving an $A_{k\geq 1}$ singularity corresponds to the function $f (x)= x_1^2 + x_2^2+x_3^2+x_4^{k+1}$~.

Locally flops can be distinguished by two invariants. The first of these is the normal bundle of the exceptional curve $C$. Laufer~\cite{Laufer:1981} showed that the only normal bundles that can occur are $\mathcal{O}_C(-1)\oplus \mathcal{O}_C(-1)$, $\mathcal{O}_C(-2)\oplus \mathcal{O}_C(0)$ and $\mathcal{O}_C(-3)\oplus \mathcal{O}_C(1)$. All flops of type $(-1,-1)$ are locally equivalent to the Atyiah flop, those of type $(-2,0)$ are equivalent to Reid's pagoda flop involving an $A_{k>1}$ singularity and all other flops are in class $(-3, 1)$. The second invariant is the length of the flop. All the flops involving $A_k$ singularities, that is those of type $(-1,-1)$ and $(-2,0)$, have length $1$, while flops involving $D_k$, $E_6$, $E_7$ and $E_8$ singularities have lengths between $2$ and $6$ (the flops with $E_6$, $E_7$ and $E_8$ singularities currently being conjectural) \cite{KatzMorrison:1992}.

Going beyond such local aspects is difficult in general. In particular, given a global description of $X^+$, it is difficult to find a global description of $X^-$. The purpose of the present work is to provide such a description in the case of flops for which $X^+$ is a Calabi-Yau threefold realised as a generic K\"ahler-favourable complete intersection in a product of projective spaces (a CICY threefold). For a definition of K\"ahler favourability see Section~\ref{sec:favourability}. We note in passing that many of the ideas presented below appear to generalise to cases where $X^+$ is a Calabi-Yau threefold realised as a complete intersection inside a toric ambient space, and this generalisation will be the subject of future work.

\vspace{11pt}
\noindent
\subsection{Results.}
Throughout this paper $X$ will be a CICY threefold. We will write the product of projective spaces in which $X$ is defined as $\mathbb{P}^n\times \vec{\mathbb{P}}$, where $\vec{\mathbb{P}}$ is itself a product of projective spaces. We will also assume that the embedding of $X$ in $\mathbb{P}^n\times \vec{\mathbb{P}}$ is K\"ahler-favourable, i.e.\ the K\"ahler cone of $X$ can be identified with that of the embedding product of projective spaces. In particular, the K\"ahler cone of $X$ will be finitely-generated. The multi-degrees of the equations defining $X$ will be collected in a configuration matrix, as reviewed in Section~\ref{sec:CICY} below.

Manifolds described by the same configuration matrix belong to the same deformation family and we will refer to two smooth members of such a deformation family as `isomorphic', which is hence strictly an isomorphism only in the category of smooth manifolds (by Wall's theorem \cite{wall}), i.e.\ a diffeomorphism.

In Section~\ref{sec:class} we give a simple criterion, proven in Prop.~\ref{prp:classification}, for determining from the configuration matrix which boundaries of the K\"ahler cone are also boundaries of the effective cone and which ones may correspond to flops. In \srefs{sec:expl_flops} and \ref{sec:expl_flops2} we show that, for generic defining polynomials, all the boundaries that are not boundaries of the effective cone do correspond to flops and we explicitly construct the flopped threefolds.
\medskip

\noindent{\bfseries Theorem.} {\itshape Let $X$ be a K\"ahler-favourable CICY threefold with generic defining equations. The configuration matrix of $X$ indicates the presence (or absence) of flops and in particular that flops occur only in the following two cases: 

\medskip

\noindent\textbf{Type 1:} The configuration matrix is of the form
\begin{equation}
\left[\begin{array}{c|ccccccc}\mathbb{P}^n&1&1&\ldots&1&0&\ldots&0\\ \vec{\mathbb{P}} &\vec{q}_1&\vec{q}_2&\ldots&\vec{q}_{n+1}&\vec{q}_{n+2}&\ldots&\vec{q}_{K}\end{array}\right]\,.
\label{eq:intro_case1}
\end{equation}
In this case, the CICY threefold can be flopped in the $\mathbb{P}^n$ direction
\begin{enumerate}
\item[(a)] to an isomorphic Calabi-Yau threefold, if $\vec{q}_1=\vec{q}_2=\ldots=\vec{q}_{n+1}$,
\item[(b)] to a typically, but not necessarily, different Calabi-Yau threefold, otherwise.
\end{enumerate}

\medskip

\noindent\textbf{Type 2:} The configuration matrix is of the form
\begin{equation}
\left[\begin{array}{c|cccccccc}
\mathbb{P}^n&2&1& \ldots & 1 & 0 & \ldots & 0 \\
\vec{\mathbb{P}} &\vec{p}_1&\vec{p}_2&\ldots&\vec{p}_n&\vec{p}_{n+1}&\ldots&\vec{p}_K\end{array}\right]\,,
\label{eq:intro_case2}
\end{equation}
including the case with no `1' entries in the first row. In this case the manifold can be flopped along the $\mathbb{P}^n$ direction to an isomorphic Calabi-Yau threefold.}
\medskip

We would like to point out that unlike the flops involving a row of Type~1, flops involving a row of Type~2 are not related to the configuration matrix splittings discussed in Refs.~\cite{Candelas:1987kf,Green:1988wa,Green:1988bp} in the context of conifold transitions.

\medskip

The proof of the above statement also shows that there are no K\"ahler cone boundaries at which a divisor shrinks and the Calabi-Yau volume remains non-vanishing. We note in passing that the flop associated to a row of Type~1 can be seen from a GLSM description, while flops of Type~2 appear to require one to go beyond a standard GLSM analysis.
 
 \medskip

Rows of the two types above are not rare, but in fact make up the vast majority of rows appearing in configuration matrices in the classification of K\"ahler-favourable CICYs \cite{Candelas:1987kf,Green:1987cr, Anderson:2017aux}. Indeed, among the 30,924 rows which appear in these configuration matrices, there are 28,531 rows of Type~1 and 2,272 rows of Type~2, totaling 30,803 rows. Hence almost every boundary of the K\"ahler cones of K\"ahler-favourable CICYs corresponds to a flop.

\medskip

\noindent{\bfseries Infinite flop chains.} When the configuration matrix contains multiple rows of Type~1(a) or 2, the CICY admits multiple flops to isomorphic manifolds, and hence infinite chains of flops. One finds this occurs for 505 of the 4874 K\"ahler-favourable CICYs. In this case, the extended K\"ahler cone consists of an infinite number of sub-cones, and when $h^{1,1}(X)>2$ it has infinitely many edges. The simplest case occurs for $h^{1,1}(X)=2$, when both boundaries of the K\"ahler cone correspond to flops to isomorphic threefolds, occurring for e.g.\

\vspace{-.3cm}
\begin{align}
\left[\begin{array}{c|ccc}\mathbb{P}^3 &2&1&1\\\mathbb{P}^3 &2&1&1\end{array}\right]\,, \quad
\left[\begin{array}{c|ccccc}\mathbb{P}^4 &1&1&1&1&1\\\mathbb{P}^4 &1&1&1&1&1\end{array}\right]\,, \quad
\left[\begin{array}{c|ccccc}\mathbb{P}^3 &0&1&1&1&1\\\mathbb{P}^5 &2&1&1&1&1\end{array}\right]\,,
\end{align}
and the envelope of the extended K\"ahler cone is an irrational two-dimensional cone - see also Refs.~\cite{Brodie:2020fiq, Brodie:2021ain, Brodie:2021nit, Brodie:2021zqq, Constantin:2022jyd}.

\medskip
In Section~\ref{sec:expl_flops} we prove the following result, which is the content of Thm.~\ref{thm_flop1} and Cor.~\ref{crl:Pic_iso} of Prop.~\ref{prop:divisors_1}.

\medskip
\noindent{\bfseries Theorem.} {\itshape For a CICY threefold flop $X \dashrightarrow X'$ on a row of Type~1 of the configuration matrix~\eqref{eq:intro_case1}, the flopped manifold can be described as a complete intersection inside a toric ambient space, namely
\be
\begin{array}{c c c c | c c c c c c}
x_1 & \ldots & x_{n+1} 	& y 					& P_1 & \ldots & P_{n+1} & P_{n+2} & \ldots & P_K\\
\hline
1 & \ldots & 1 & 0	 						& 1 & \ldots & 1 & 0 &  \ldots & 0 \\
-\vec{q}_1 & \ldots & -\vec{q}_{n+1} & \square 	& 0 & \ldots & 0 & \vec{q}_{n+2} & \ldots & \vec{q}_K 
\end{array}
\label{eq:flp_wgts}
\ee
where $y$ represents collectively the coordinates on the product of projective spaces $\vec{\mathbb{P}}$ appearing in the second row of the configuration matrix \eqref{eq:intro_case1} while $\square$ stands for their unchanged weights. The isomorphism relating the Picard groups of the original and flopped Calabi-Yau threefolds $X$ and $\fl{X}$ can be written as
\be
\mathrm{Pic}(X) \to \mathrm{Pic}(\fl{X}) \, \colon
\hspace{.2cm}
\vec{v} \, \mapsto
\left( \begin{array}{c c}
-1 & \vec{0}^{\,\mathrm{T}} \\[4pt]
\sum_{k = 1}^{n+1} \vec{q}_k  & \mathbbm{1}
\end{array} \right)
\vec{v} \,,
\ee
where divisor classes on $X$ are written with respect to the basis of (the pullbacks to $X$ of) the hyperplane classes of the projective space factors in the embedding space, and on $\fl{X}$ with respect to the basis of (the pullbacks to $\fl{X}$ of) the classes corresponding to the rows of the weight system in \eref{eq:flp_wgts}.} 
\medskip

Note that the K\"ahler cone of the flopped manifold may in general be difficult to obtain.
\medskip

\noindent{\bfseries Counting the number of collapsing curves (rows of Type 1).} 
In Section~\ref{sec:expl_flops_theflop} we show that the singular threefold involved in a CICY flop on a row of Type~1 of the configuration matrix~\eqref{eq:intro_case1} can be deformed to a smooth CICY threefold $\defm{X}$ with configuration matrix
\be
\big[\!\!
\begin{array}{c | c c c c}
\vec{\mathbb{P}} & \sum_{j=1}^{n+1} \vec{q}_j & \vec{q}_{n+2} & \hdots & \vec{q}_K \\
\end{array}
\!\!\big]
\ee
to complete a conifold transition, as in the well-known story of splittings in Refs.~\cite{Candelas:1987kf,Green:1988wa,Green:1988bp}. The contracting curves are all in the same curve class, and their number $\mathrm{num}\big(\mathbb{P}^1\big)$, corresponding to the genus-zero Gromov-Witten invariant, is determined by the difference in Euler characteristics between the original CICY $X$ and the CICY $\defm{X}$ on the deformation branch,
\be
\mathrm{num}\big(\mathbb{P}^1\big) = \tfrac{1}{2}\big(\ec(X) - \ec(\defm{X})\big) \,.
\ee
\medskip

In Section~\ref{sec:expl_flops2} we prove the following result, which is the content of Thm.~\ref{thm:flops2} and Cor.~\ref{crl:picard_iso2} of Prop.~\ref{prop:divisors_2}.

\medskip

\noindent{\bfseries Theorem.} {\itshape
For a CICY threefold flop $X \dashrightarrow X'$ on a row of Type~2 of the configuration matrix~\eqref{eq:intro_case2}, the flopped Calabi-Yau threefold $X'$ is always isomorphic to the threefold $X$. The  isomorphism induced on the Picard group of $X$ by the flop is
\be
\mathrm{Pic}(X) \to \mathrm{Pic}(X) \, \colon
\hspace{.2cm}
\vec{v} \, \mapsto
\left( \begin{array}{c c}
-1 & 0 \\[4pt]
\vec{q}_1 + 2\sum_{k=2}^n \vec{q}_k  & 1 
\end{array} \right)
\vec{v} \,,
\ee
where divisor classes on $X$ are written with respect to the basis of (the pullbacks to $X$ of) the hyperplane classes of the projective space factors in the embedding space. 
The K\"ahler cone of the flopped manifold is generated by
\be
\Big\langle \big(-1,2\big(\sum_{k=1}^n \vec{q}_k \big) -\vec{q}_1 \big) \,,\, (0,1,0,\ldots,0) \,,\, \ldots \,,\, (0, \ldots , 0 , 1) \big) \Big\rangle \,.
\ee
The singular manifold at the intermediate point of the flop can be deformed to a complete intersection in a toric ambient space described as 
\be
\begin{array}{c c | c c c c}
\xi  						& y			& & & & \\
\hline \rule{0pt}{2.5ex}
\sum_{i=1}^n \vec{q}_i	& \square 	& 2\sum_{i=1}^n \vec{q}_i & \vec{q}_{n+1} & \hdots & \vec{q}_K  \\
\end{array}
\ee
to complete a conifold transition to a Calabi-Yau $\defm{X}$, where $y$ represents collectively the coordinates on the product of projective spaces $\vec{\mathbb{P}}$ in the second row of the configuration matrix \eqref{eq:intro_case2} while $\square$ stands for their unchanged weights. 
}
\medskip

\noindent{\bfseries Counting the number of collapsing curves (rows of Type 2).} 
In contrast to the case of flops on rows of Type~1, in the case of flops on rows of Type~2 some curves are in a class $[C]$ while others are in the class~$2[C]$. In Section~\ref{sec:counting_curves_type2} we show that the numbers $\mathrm{num}\big(\mathbb{P}^1_{[C]}\big)$ and $\mathrm{num}\big(\mathbb{P}^1_{2[C]}\big)$ of contracting curves in each class are given by the two relations
\be
\begin{gathered}
\mathrm{num}\big(\mathbb{P}^1_{[C]}\big) + \mathrm{num}\big(\mathbb{P}^1_{2[C]}\big) = \tfrac{1}{2}\big(\ec(X) - \ec(\defm{X})\big)
\,, \\
\mathrm{num}\big(\mathbb{P}^1_{[C]}\big) + 2 \,\mathrm{num}\big(\mathbb{P}^1_{2[C]} \big) = \tfrac{1}{2} \, c_2(X) \cdot \Big( -2 \,,\, \vec{q}_1 + 2\sum_{k=2}^n \vec{q}_k \Big) \,.
\end{gathered}
\ee
Note that $\mathrm{num}\big(\mathbb{P}^1_{[C]}\big)$, $\mathrm{num}\big(\mathbb{P}^1_{2[C]}\big)$ are genus-zero Gromov-Witten invariants of the original CICY \cite{Wilson1997FlopsTI}.

\vspace{12pt}
{\bfseries Acknowledgements.} We are grateful to Lara Anderson, Antonella Grassi, and James Gray for useful discussions. The work of CB has been supported by the NSF grant PHY-2014086 and the John Templeton Foundation grant 61149. The work of AC is supported by the EPSRC grant EP/T016280/1. The work of FR is supported by startup funding from Northeastern University.

\section{Background}

\subsection{Flops for Calabi-Yau threefolds} 

Formally, a simple flop between two smooth Calabi-Yau threefolds $X^+$ and $X^-$ is a birational map which is an isomorphism away from smooth rational curves $C^+$ and $C^-$ on $X^+$ and $X^-$ respectively. The curves $C^+$ and $C^-$ can be contracted to points in $X^+$ and $X^-$, respectively, to produce the same singular threefold $X_{\rm sing}$. What distinguishes a flop from, say, the identity map, is the condition that $X^+$ contains a divisor $D^+$ with a positive intersection $D^+\cdot C^+>0$ such that the corresponding divisor $D^-$ on $X^-$ under the identification $X^+- C^+\cong X^-- C^-$ has a negative intersection $D^-\cdot C^-<0$. The divisor $D^+$ is not unique. This is summarised in the following global characterisation of a flop between two smooth Calabi-Yau threefolds (see e.g.\ Ref.~\cite{KollarMori:1998} for a general definition of a flop between two normal varieties).

\begin{defn}\label{def:flop}
Let $X^+$ and $X^-$ be smooth Calabi-Yau threefolds and $$\pi^+ : X^+ \to X_\mathrm{sing}~,\qquad \pi^- : X^- \to X_\mathrm{sing}~,$$ be small contractions, mapping a finite number of curves $\mathcal{C}_i^+$ on $X^+$ and $\mathcal{C}_i^-$ on $X^-$ to the same singular points $p_i\in X_{\rm sing}$ and being isomorphisms away from these curves. The map $(\pi^-)^{-1}\circ\pi^+:X^+-\{\mathcal{C}^+_i\} \rightarrow X^--\{\mathcal{C}^-_i\}$ is an isomorphism and so defines a birational map $$ \phi : X^+ \dashrightarrow X^-~.$$ Then $\phi$ is a flop if there exists a divisor $D^+$ on $X^+$ such that for every $i$, $D^+\cdot \mathcal{C}_i^+>0$ and $D^-\cdot \mathcal{C}_i^-<0$, where $D^-$ is the divisor on $X^-$ to which $D^+$ is mapped.
\label{def:flop_def}
\end{defn}

This definition implies the following commutative diagram: 
\vspace{-.2cm}
\begin{equation}\label{flop:comm_diag}
\begin{aligned}
\begin{tikzpicture}
\node (X) at (0,0) {$X^-$};
\node (Xt) at (3,0) {$X^+$};
\node (Xs) at (1.5,-1.3) {$\;X_{\rm sing}$};
\node (pim) at (.4,-.7) {$\pi^-$};
\node (pip) at (2.6,-.7) {$\pi^+$};
\draw[<->,dashed] (X)+(.4,0) -- (Xt) 
	node[midway, above] {$\;\phi$};
\draw[->,] (X)+(.3,-0.2) -- (Xs);
\draw[->,] (Xt) -- (Xs); 
\end{tikzpicture}
\end{aligned}
\end{equation}
\vspace{-.4cm}

In general $X^+$ and $X^-$ are topologically distinct. Since a flop is an isomorphism in codimension one, if $X^+$ has trivial canonical divisor class, so too will $X^-$, and hence the Calabi-Yau condition is preserved under a flop. Moreover, $X^-$ and $X^+$ share the same Hodge numbers. However, finer topological invariants, such as the intersection numbers and the second Chern class of the manifold, do change under a flop. Concretely, if $D^+$ is a divisor on $X^+$ and $D^-$ the corresponding divisor on $X^-$, the triple self-intersection form and the $c_2$-form on $H^2(X^+,\mathbb Z)\cong H^2(X^-,\mathbb Z)$ change in the following way: 
\begin{equation}
(D^-)^3 = (D^+)^3 - \sum_{i=1}^{N} (D^+\cdot \mathcal{C}^+_i)^3~,
\label{eq:dijk_change}
\end{equation}
\vspace{-.7cm}
\begin{equation}
c_2(X^-)\cdot D^- = c_2(X^+)\cdot D^+ +2 \sum_{i=1}^{N} D^+\cdot \mathcal{C}^+_{i}~,
\label{eq:c2_change}
\end{equation}
where $\mathcal{C}^+_{1},\mathcal{C}^+_{2},\ldots,\mathcal{C}^+_{N}$ are the isolated exceptional $\mathbb{P}^1$ curves contracted in the flop. In general, these curves either belong to a single primitive homology class $[C]\in H_2(X^+,\mathbb Z)$ or some belong to $[C]$ and some to $2[C]$. We will denote by $n_1\leq N$ the number of curves in the class $[C]$ and by $n_2 \leq N$ the number in $2[C]$. The class $[C]$ is perpendicular, with respect to the intersection form on $X^+$, to a codimension one face of the K\"ahler cone $\mathcal{K}(X^+)$. This face is a wall separating $\mathcal{K}(X^+)$ from the image of the K\"ahler cone $\mathcal{K}(X^-)$ of the flopped manifold under the identification $H^2(X^+,\mathbb R)\cong H^2(X^-,\mathbb R)$. 

\subsection{CICYs and configuration matrices}
\label{sec:CICY}
A Complete-Intersection Calabi-Yau (CICY) manifold is constructed as the common vanishing locus of a number of polynomials inside a product of complex projective spaces. Conventionally, a CICY is represented by a configuration matrix,
\be
X \sim
\left[
\begin{array}{c | c c c }
\mathbb{P}^{n_1} & q_1^1 & \hdots & q_K^1\\
\vdots & \vdots & \ddots & \vdots \\
\mathbb{P}^{n_m} & q_1^m & \hdots & q_K^m\\
\end{array}
\right] \,.
\label{eq:conf_mat_gen}
\ee
Here the ambient space is $\mathcal{A} = \mathbb{P}^{n_1} \times \ldots \times \mathbb{P}^{n_m}$, and each column describes one of the defining polynomials $\{P_1 \,, \ldots , P_K\}$. As such, the entry $q_j^i$ represents the degree of the polynomial $P_j$ with respect to the coordinates of the projective space $\mathbb{P}^{n_i}$. A configuration matrix in fact describes a family of manifolds, corresponding to the different possible choices of the defining polynomials. In the below, we will always consider generic choices, which are known to give smooth manifolds by Bertini's Theorem. However, we note that special choices of the defining polynomials can result in smooth manifolds whose geometrical properties are qualitatively different from the generic case. (We will see an example of this at the end of Section~\ref{sec:class}.)

The dimension of the CICY is equal to the dimension of the ambient space minus the number of polynomials, $\dim(X)=\sum_{i=1}^m n_i-K$,  and for most of the present discussion ${\rm dim}(X)=3$. The requirement that the manifold be Calabi-Yau corresponds to the condition that the entries in each row should add up to one plus the dimension of the corresponding projective space, i.e.\ $\sum_{j=1}^K q_i^j = n_i+1$. In the simplest examples, there is only one projective space in the ambient space and only one defining polynomial, as in
\be
\left[\!\!
\begin{array}{c | c }
\mathbb{P}^2 \!\!\!&\! 3
\end{array}
\!\!\right] \,, \quad 
\left[\!\!
\begin{array}{c | c }
\mathbb{P}^3  \!\!\!&\!  4
\end{array}
\!\!\right] \,, \quad 
\left[\!\!
\begin{array}{c | c }
\mathbb{P}^4  \!\!\!&\!  5
\end{array}
\!\!\right] \,,
\ee
which respectively describe an elliptic curve, a K3 surface, and a Calabi-Yau threefold. Below we will make use of the following CICY threefolds, taken from the CICY list~\cite{Candelas:1987kf,Green:1987cr}, which contains $7890$ configuration matrices:
\be
\begin{aligned}
X_{7858}
&~\sim\,
\left[
\begin{array}{c | c c }
\mathbb{P}^1[x] & 1 & 1 \\
\mathbb{P}^4[y] & 3 & 2 \\
\end{array}
\right]
\hspace{-6pt}
&:~~\qquad&
\left\{ \begin{array}{c} F_1^{(3)}(y)\,x_1 + F_2^{(3)}(y)\,x_2 = 0 \\ G_1^{(2)}(y)\,x_1 + G_2^{(2)}(y)\,x_2 = 0 \end{array} \right\} \,, \\ 
X_{7887}
&~\sim\,
\left[
\begin{array}{c | c }
\mathbb{P}^1[x] & 2 \\
\mathbb{P}^3[y] & 4 \\
\end{array}
\right]
\hspace{-6pt}
&:~~\qquad&
\left\{ \begin{array}{c} H_1^{(4)}(y)\,x_1^2 + H_2^{(4)}(y)\,x_1x_2 + H_3^{(4)}(y)\,x_2^2 = 0  \end{array} \right\} \,, \\ 
X_{7833}
&~\sim\,
\left[
\begin{array}{c | c c }
\mathbb{P}^2[x] & 2 & 1 \\
\mathbb{P}^3[y] & 1 & 3 \\
\end{array}
\right]
\hspace{-6pt}
&:~~\qquad&
\left\{ \begin{array}{r@{\;}l} Q_1^{(1)}(y)\,x_1^2 +
Q_2^{(1)}(y)\,x_1x_2 + \ldots + Q_6^{(1)}(y)\,x_3^2&= 0 \\ R_1^{(3)}(y)\,x_1 + R_2^{(3)}(y)\,x_2 + R_3^{(3)}(y)\,x_3&= 0 \end{array} \right\} \,.
\end{aligned}
\ee
In these configuration matrices we have written explicitly the variable names for the coordinates in the projective spaces. We also included in brackets the explicit defining equations, in which the superscripts on the functions indicate their degrees, while the subscripts are mere labels.

\medskip

\subsection{Favourable embeddings and K\"ahler favourability}\label{sec:favourability}
The Picard group of a projective space is generated by a single element, the hyperplane class $H$, thus $\mathrm{Pic}(\mathbb{P}^{n_i}) = \langle H_i \rangle$. Hence, the Picard group of the ambient space $\mathcal{A}$ is generated as
\be
\mathrm{Pic}(\mathcal{A}) = \langle H_1 \,,\, \ldots \,,\, H_m \rangle \,,
\ee
where by abuse of notation $H_i$ now represents the pullbacks of the hyperplane class of $\mathbb{P}^{n_i}$ to $\mathcal{A}$ by the projection map. Given a CICY manifold $X\subseteq \mathcal{A}$, the $H_i$ restrict to divisor classes on $X$, which we write as~$D_i$. If $\mathrm{Pic}(X) \otimes \mathbb{R} = \langle D_1 \,,\, \ldots \,,\, D_m \rangle \otimes \mathbb{R}$, so that $h^{1,1}(X) = h^{1,1}(\mathcal{A})=m$, the embedding is said to be {\itshape favourable}. If, in addition, the K\"ahler cone of $X$ can be identified with the K\"ahler cone of $\mathcal{A}$,  we say that the description is {\itshape K\"ahler-favourable}. In this case the K\"ahler cone of $X$, denoted by $\mathcal K(X)$, corresponds to the positive quadrant in $\mathrm{Pic}(X) \otimes \mathbb{R}$ with respect to the basis $\{D_i\}$. In the below, we will only consider K\"ahler-favourable embeddings. Note that the vast majority of CICY threefolds can be favourably embedded  \cite{Anderson:2017aux}. 

With respect to the basis $\{D_i\}$, the triple intersection numbers of $X$ are denoted by $d_{ijk}=D_i\cdot D_j\cdot D_k$. The dual set of curve classes will be denoted by $\{C_1,\ldots C_m\}$, such that $D_i\cdot C_j = \delta_{i,j}$. 

\section{Classification of K\"ahler cone boundaries for CICYs}
\label{sec:class}

Let $X$ be a K\"ahler-favourable CICY threefold, so that the K\"ahler cone $\mathcal{K}(X)$ is the positive quadrant in the basis of divisor classes $\{D_i\}$ that descend from the ambient projective spaces. Let $\bnd_i$  denote the boundary of the K\"ahler cone such that 
$
\bnd_i = \langle D_1, \ldots D_{i-1},D_{i+1},\ldots D_m \rangle~
$
is dual to the curve class $C_i$, that is $\bnd_i  = (C_i)^\perp$. This gives a one-to-one correspondence between the $i^\mathrm{th}$ row of the configuration matrix, the boundary $\bnd_i$, and the curve class $C_i$.
Each boundary $\bnd_i$ then falls into one of the three categories:
\begin{enumerate}
\item[(1)] {\itshape Flop wall.} As the K\"ahler form on $X$ approaches a flop wall, the volumes of the curves in $C_i$ go to zero, while the volumes of all divisors and the volume of $X$ remain non-zero. 
\item[(2)] {\itshape Zariski wall.} As the K\"ahler form on $X$ approaches a Zariski wall, the volume of a divisor $D$ in $X$ goes to zero, either by collapsing to a curve or to a point, while the volume of $X$ remains non-zero. 
\item[(3)]{\itshape Effective cone boundary.} As the K\"ahler form on $X$ approaches an effective cone boundary, the volume of $X$ goes to zero. Indeed, for any divisor class $D$ lying on such a boundary, $D^3=0$.
\end{enumerate}

Our purpose in this section is to give a simple criterion for determining from the configuration matrix which boundaries of the K\"ahler cone may correspond to flops.

\begin{prp}\label{prp:classification} Let $X$ be a K\"ahler-favourable CICY threefold. The configuration matrix of $X$ indicates the possibility of flop transitions and in particular that flops can only occur in the following two cases: 

\medskip

\noindent\textbf{Type 1:} The configuration matrix is of the form
\begin{equation}
\left[\begin{array}{c|ccccccc}\mathbb{P}^n&1&1&\ldots&1&0&\ldots&0\\ \vec{\mathbb{P}} &\vec{q}_1&\vec{q}_2&\ldots&\vec{q}_{n+1}&\vec{q}_{n+2}&\ldots&\vec{q}_{K}\end{array}\right] \,.
\end{equation}

\medskip

\noindent\textbf{Type 2:} The configuration matrix is of the form
\begin{equation}
\left[\begin{array}{c|cccccccc}
\mathbb{P}^n&2&1& \ldots & 1 & 0 & \ldots & 0 \\
\vec{\mathbb{P}} &\vec{p}_1&\vec{p}_2&\ldots&\vec{p}_n&\vec{p}_{n+1}&\ldots&\vec{p}_K\end{array}\right]\,,
\end{equation}
including the case with no `1' entries in the first row.
\end{prp}

\begin{proof}
We begin by determining which K\"ahler cone walls are not effective cone boundaries. If $\bnd_i$ is a boundary of the effective cone, the cubic form vanishes on every $D\in\bnd_i$, that is $D^3=0$. In particular, from the definition of $\bnd_i$, this implies the following condition on the triple intersection numbers:
\be
d_{jkl} \equiv D_j \cdot D_k \cdot D_l = 0, \qquad \forall \, j,k,l \neq i ~.
\label{eq:trip_int_cond}
\ee
We will now derive the implications of these conditions for the CICY. The triple intersection numbers follow straightforwardly from the configuration matrix and are given by the ambient space intersection
\be
d_{jkl} = H_j \cdot H_k \cdot H_l \cdot \prod_{s=1}^K \left( \sum_{r=1}^m q_s^r H_r \right) \,,
\label{eq:trip_int_gen}
\ee
which restricts the intersection $H_j \cdot H_k \cdot H_l$ to $X$ by further intersecting with the divisor classes $\sum_{r=1}^m q_s^r H_r$ which correspond to the defining polynomials. Since the number of defining polynomials is $K = n_1+\ldots + n_m -3$, the intersection product in \eref{eq:trip_int_gen} contains only terms with $n_1+\ldots + n_m$ factors and the only non-zero contributions come from terms of the form 
\begin{equation*}
H_1^{n_1}\cdot H_2^{n_2}\cdot \ldots\cdot H_m^{n_m}~.
\end{equation*}
For intersection numbers $d_{jkl}$ with $j,k,l\neq i$ the contribution $H_i^{n_i}$ can only come from the polynomial part of \eref{eq:trip_int_gen}. Hence, a non-zero $d_{jkl}$ with $j,k,l\neq i$ requires at least $n_i$ polynomials which depend on the coordinates of the $i^{\rm th}$ projective space. This happens if the $i^{\rm th}$ row is, up to permutations of columns, of one of the following two types: 
\be
\begin{aligned}
\mathrm{Type}~1:&~~
\big[
\begin{array}{c} \mathbb{P}^{n} \end{array}
\big|
\underbrace{\begin{array}{c c c}1 & \ldots & 1\end{array}}_{n+1}
\begin{array}{c c c}0 & \ldots & 0\end{array}
\big] \,,
\\
\mathrm{Type}~2:&~~
\big[
\begin{array}{c} \mathbb{P}^{n} \end{array}
\big|
\begin{array}{c} 2 \end{array}
\underbrace{\begin{array}{c c c} 1 & \ldots & 1\end{array}}_{n-1}
\begin{array}{c c c}0 & \ldots & 0\end{array}
\big] \,.
\end{aligned}
\label{eq:two_cases}
\ee
Conversely, if the $i^{\rm th}$ row is not of this form, then the conditions \eqref{eq:trip_int_cond} are satisfied since the number of  $H_1$ factors in \eref{eq:trip_int_gen} is always less than $n_1$. In this case the boundary $\bnd_i$ is a boundary of the effective cone. Thus only the above two types of rows can lead to K\"ahler cone boundaries which correspond to flops or Zariski walls.
\end{proof}

 In \srefs{sec:expl_flops} and \ref{sec:expl_flops2} we will show that, for generic defining polynomials, all of these boundaries in fact correspond to flops. In particular, this implies that the K\"ahler cone of a generic K\"ahler-favourable CICY never has a Zariski wall.
For non-generic choices of the defining polynomials, it can happen that the flopping $\mathbb{P}^1$s coalesce into a divisor, which then contracts to a curve, the flop wall becoming a Zariski wall. An explicit example of this is given in Section~4.1 of Ref.~\cite{Brodie:2021nit}, while some general analysis can be found in Section~3.3 of Ref.~\cite{katz1996}. We leave a systematic treatment of this phenomenon in the case of CICYs to future work.

\section{Performing flops on rows of Type 1}
\label{sec:expl_flops}

A configuration matrix containing a row of Type~1 in the classification of \sref{sec:class} is, up to permutations of rows and columns, of the form
\be
X \sim 
\left[
\begin{array}{c | c c c c c c}
\mathbb{P}^n[x] & 1 & \ldots & 1 & 0 & \ldots & 0 \\
\vec{\mathbb{P}}[y] & \vec{q}_1 & \ldots & \vec{q}_{n+1} & \vec{q}_{n+2} & \hdots & \vec{q}_K \\
\end{array}
\right]
\,.
\label{eq:conf_gen_111}
\ee
Here $x=[x_1:x_2:\ldots:x_{n+1}]$ stands for the coordinates on the first projective space of dimension $n$ while $\vec{\mathbb{P}}[y]$ represents the product of the remaining projective spaces with coordinates collectively denoted by $y$.
Hence, the ambient space is of the form $\mathcal{A}=\mathbb{P}^n\times \vec{\mathbb{P}}$. The vector $\vec{q}_i$ collects the entries in the $i^{\rm th}$ column of the configuration matrix after dropping the first row.

Below, we will denote the CICY and the associated flopped Calabi-Yau by $X$ and $\fl{X}$, respectively, instead of the symbols $X^+$ and $X^-$ from Definition~\eqref{def:flop_def}, to notationally differentiate the CICY from the flopped manifold, the latter of which will not generically have a CICY description.

\subsection{The contraction map} 
\label{sec:expl_flops_theflop}
We show that for generic defining equations the threefold \eqref{eq:conf_gen_111} admits a flop along the $\mathbb P^n$ direction. We do this by explicitly constructing the contraction map.  

\begin{prp}
A generic K\"ahler-favourable CICY threefold whose configuration matrix contains a row of Type 1 admits a small contraction $\pi:X\rightarrow X_{\rm sing}$, where $X_{\rm sing}$ has isolated singularities. 
\end{prp}

\begin{proof}
The last $K-n-1$ equations in the configuration matrix in \eref{eq:conf_gen_111} define a complete intersection $Y$ inside $\vec{\mathbb{P}}[y]$,
\be
Y \sim 
\left[
\begin{array}{c | c c c }
\vec{\mathbb{P}}[y] & \vec{q}_{n+2} & \hdots & \vec{q}_K \\
\end{array}
\right] \;,
\ee
(unless $K=n+1$, in which case simply $Y = \vec{\mathbb{P}}[y]$) which one can check is a fourfold. The first $n+1$ equations of the configuration matrix are linear equations in $\mathbb{P}^n[x]$, with coefficients varying over $Y$. Generically they have no common solution. However, over a codimension one locus of $Y$ the rank of this linear system drops by one, leading to a single solution in $\mathbb{P}^n[x]$. Hence the CICY $X$ is almost everywhere isomorphic to a hypersurface inside the manifold $Y$. However over a codimension-four locus of $Y$, i.e.\ at a set of points, the rank of the linear system drops by two, and the linear system determines an entire $\mathbb{P}^1$ locus inside $\mathbb{P}^n[x]$. Hence, $X$ is the small resolution of a hypersurface in~$Y$ at a number of singular points.

To describe this more explicitly, we write out the first $n+1$ equations of the configuration matrix,
\be
F_{1,1}^{(\vec{q}_1)}(y)\,x_1 + \ldots + F_{1,n+1}^{(\vec{q}_1)}(y)\,x_{n+1} = 0 ~\,,~ \ldots ~\,,~  F_{n+1,1}^{(\vec{q}_{n+1})}(y)\,x_1 + \ldots + F_{n+1,n+1}^{(\vec{q}_{n+1})}(y)\,x_{n+1} = 0 \,.
\ee
Here the superscripts denote the multi-degree of the polynomials in $y$ with respect to the projective spaces in the product $\vec{\mathbb{P}}[y]$. It will be useful for the subsequent discussion to write these equations in matrix vector form as
\be
\vec{0} = 
F(y)\, \vec{x} \equiv
\left(
\begin{array}{c c c}
F_{1,1}^{(\vec{q}_1)}(y) & \ldots & F_{1,n+1}^{(\vec{q}_1)}(y) \\
\vdots & \ddots & \vdots \\
F_{n+1,1}^{(\vec{q}_{n+1})}(y) & \ldots & F_{n+1,n+1}^{(\vec{q}_{n+1})}(y) \\
\end{array}
\right)
\left(
\begin{array}{c} x_1 \\ \vdots \\ x_{n+1} \end{array}
\right)
\,.
\label{eq:111_mat_eq}
\ee
Since the trivial solution does not correspond to an element in $\mathbb{P}^n[x]$, the determinant of $F(y)$ must vanish, which gives a hypersurface in $\vec{\mathbb{P}}[y]$. Together with the other equations in the configuration matrix, specified by $\vec{q}_{n+2}\,, \ldots \,, \vec{q}_K$, this determines the image of $X$ under the projection $\mathbb{P}^n[x] \times \vec{\mathbb{P}}[y] \to \vec{\mathbb{P}}[y]$, which we write as
\be
\contr{X} \sim \big[\!\!
\begin{array}{c | c c c c}
\vec{\mathbb{P}}[y] & \left( {\rm det}\,F(y)=0\right) &\vec{q}_{n+2} & \hdots & \vec{q}_K \\
\end{array}
\!\!\big]\subset Y
\,.
\label{eq:111_null1}
\ee
This is a singular threefold which belongs to the family 
\be
X^{\flat} \sim \big[\!\!
\begin{array}{c | c c c c}
\vec{\mathbb{P}}[y] & \sum_{j=1}^{n+1} \vec{q}_j & \vec{q}_{n+2} & \hdots & \vec{q}_K \\
\end{array}
\!\!\big]
\,.
\label{eq:111_null11}
\ee
At the locus in $\vec{\mathbb{P}}[y]$ where the rank of $F(y)$ further drops to $n-1$, the linear system~\eqref{eq:111_mat_eq} determines an entire $\mathbb{P}^1$ inside $\mathbb{P}^n[x]$. This implies that the threefold $X$ is a small resolution of the singular threefold $\contr{X}$ at a set $S$ of points  given by 
\be
S \sim 
\big[\!\!
\begin{array}{c | c c c c}
\vec{\mathbb{P}}[y] &
\left( \mathrm{rank}\, F(y) \leq n-1\right)
& \vec{q}_{n+2} & \hdots & \vec{q}_K \\
\end{array}
\!\!\big]
\,.
\label{eq:111_null2}
\ee
(That this locus is indeed a set of points follows from general properties of determinantal varieties - see for example Section~2 of Ref.~\cite{Jockers:2012zr} - by which the locus over which the rank of the matrix $F(y)$ drops by two is codimension four.) Conversely, this gives rise to the small contraction map
\be
\pi: X \rightarrow X_{\rm sing}~.
\ee
Contracting each of these~$\mathbb{P}^1$-curves to a point produces the singular threefold $\contr{X}$. 
\end{proof}

\begin{rmk}
The threefold $X$ in \eref{eq:conf_gen_111} is on the resolution branch of the conifold transition \cite{Candelas:1989js}
\be
\begin{array}{c c c c c}
\defm{X} & \longleftrightarrow & \contr{X} & \longleftrightarrow & X
\end{array}
\ee
Hence, when we construct the flopped space below, we will be constructing the other possible resolution branch $\fl{X}$ of this conifold transition, schematically
\vspace{-.1cm}
\be
\begin{array}{c c c c c}
 		&  					& 			&  			& X
\vspace{-.1cm} \\ 
 		&  					& 			& \reflectbox{\rotatebox[origin=c]{-25}{$\longleftrightarrow$}}	&	\\
\defm{X} 	& \longleftrightarrow 	& \contr{X} 	&  			& 			\\
 		&  					& 			& \reflectbox{\rotatebox[origin=c]{25}{$\longleftrightarrow$}}	&
\vspace{-.1cm} \\
 		&  					& 			&  			& \fl{X}	\\
\end{array}
\ee
\end{rmk}

The Calabi-Yau threefold $X$ contracts to the singular variety $\contr{X}$ upon collapsing a number of $\mathbb{P}^1$ curves to points. This is one half of a flop transition. The Calabi-Yau threefold $\fl{X}$ on the other side of the flop also admits a contraction of $\mathbb{P}^1$s to the same singular variety $X_\mathrm{sing}$. Below we provide a general algorithm for  constructing the flopped threefold $\fl{X}$ explicitly. However, it is useful to start with an example.

\begin{ex}
Consider a generic Calabi-Yau threefold $X$ with configuration matrix and defining equations given by
\be
X=X_{7858}
~\sim\,
\left[
\begin{array}{c | c c }
\mathbb{P}^1[x] & 1 & 1 \\
\mathbb{P}^4[y] & 3 & 2 \\
\end{array}
\right]
\,:~~\qquad
\left\{ \begin{array}{c} P_1(x,y)= F_{1,1}^{(3)}(y)\,x_1 + F_{1,2}^{(3)}(y)\,x_2 = 0 \\  P_2(x,y)= F_{2,1}^{(2)}(y)\,x_1 + F_{2,2}^{(2)}(y)\,x_2 = 0 \end{array} \right\} \,. \\ 
\label{eq:cicy_7858}
\ee
The threefold $X$ is the small resolution of $X_\mathrm{sing}$ at a set of points $S$, where
\be
\begin{aligned}
X_\mathrm{sing} &= \left\{ y \in \mathbb{P}^4[y] \,\big{|}\, F_{1,1}^{(3)}\!(y)\,F_{2,2}^{(2)}\!(y) = F_{1,2}^{(3)}\!(y)\,F_{2,1}^{(2)}\!(y) \right\} \,,\\  
S &= \left\{ y \in \mathbb{P}^4[y] \, \big{|} \, F_{1,1}^{(3)}\!(y)= F_{2,2}^{(2)}\!(y) = F_{1,2}^{(3)}\!(y) = F_{2,1}^{(2)}\!(y) = 0\right\} \,.
\end{aligned}
\ee
Locally, the singularities of $\contr{X}$ are of the form $xy-zt=0$, so they are of the same type as in the Atiyah flop. The manifold $\contr{X}$ is a singular quintic threefold, hence the smoothed manifold $\defm{X}$ has configuration matrix
\be
\defm{X} \sim
\left[\!\!
\begin{array}{c | c c }
\mathbb{P}^4[y] & 5 \\
\end{array}
\!\!\right] \,,
\ee
and $X$ is on the resolution branch of a conifold transition from the quintic Calabi-Yau. As in the Atiyah flop, the normal bundle of the contracting curves $C$ in $X$ is
\be
N_{C,X} = \mathcal{O}_C(-1) \oplus \mathcal{O}_C(-1) \,.
\ee

By analogy with the Atiyah flop it is clear that the flopped manifold should be obtained by swapping the roles of $F_{2,1}^{(2)}(y)$ and $F_{1,2}^{(3)}(y)$, giving rise to a new manifold $\fl{X}$ defined by the following two equations:
\be
0 = F(y)^{\mathrm{T}}\, \vec{\fl{x}} =
\left(
\begin{array}{c c}
F_{1,1}^{(3)}\!(y)\, & F_{2,1}^{(2)}\!(y) \\
F_{1,2}^{(3)}\!(y)\, & F_{2,2}^{(2)}\!(y)
\end{array}
\right) 
\left(
\begin{array}{c} \fl{x}_1 \\ \fl{x}_2 \end{array}
\right)
\,\sim\,
\left\{ \begin{array}{c}
\fl{P}_1(\fl{x},y)=F_{1,1}^{(3)}\!(y)\,\fl{x}_1+ F_{2,1}^{(2)}\!(y)\,\fl{x}_2 = 0 \\
\fl{P}_2(\fl{x},y)= F_{1,2}^{(3)}\!(y)\,\fl{x}_1 + F_{2,2}^{(2)}\!(y)\,\fl{x}_2 = 0 \end{array} \right\} \,.
\ee
These equations are no longer consistent with the original projective scalings, but rather determine a complete intersection in an ambient toric variety, with a weight system and weights for the defining equations given by
\be
\fl{X} \sim 
\begin{array}{c c c c c| c c }
\fl{x}_1  & \fl{x}_2 	& y_1 & \ldots & y_5 ~ 	& \fl{P}_1  & \fl{P}_2\\
\hline
~1  & ~1 & 0	  & \ldots & 0						& 1  & 1  \\
\!\!\!-3 & \!\!\!-2 & 1 & \ldots & 1					& 0  & 0 
\end{array} \,.
\ee
The above discussion for $X$ can be repeated for this new manifold. This shows that $\fl{X}$ is a resolution of the same singular manifold $\contr{X}$ at the same set of points $S$. We will show below that $\fl{X}$ is indeed the result of flopping $X$.
\end{ex}

\subsection{The flopped manifold}\label{sec:flopped_manifold_type1}
In the above example, the equations defining $\fl{X}$ were obtained by taking the transpose operation on the defining matrix of polynomials. We now generalise this observation. 

\begin{thm}\label{thm_flop1}
A generic CICY threefold with configuration matrix~\eqref{eq:conf_gen_111} flops along the $\mathbb P^n$ direction to a complete intersection in a toric variety with a weight system and weights for the defining equations given by
\be
\fl{X} \sim 
\begin{array}{c c c c | c c c c c c}
\fl{x}_1 & \ldots & \fl{x}_{n+1} 	& y 				& \fl{P}_1 & \ldots & \fl{P}_{n+1} & P_{n+2} & \ldots & P_K\\
\hline
1 & \ldots & 1 & 0	 						& 1 & \ldots & 1 & 0 &  \ldots & 0 \\
-\vec{q}_1 & \ldots & -\vec{q}_{n+1} & \square 	& 0 & \ldots & 0 & \vec{q}_{n+2} & \ldots & \vec{q}_K 
\end{array} \,.
\label{eq:111_flp_wgts}
\ee
\end{thm}

\begin{proof}
Taking the transpose in \eref{eq:111_mat_eq} gives
\be
\vec{0} = 
F^{\mathrm{T}}(y)\,\vec{\fl{x}} \equiv
\left(
\begin{array}{c c c}
F_{1,1}^{(\vec{q}_1)}(y) & \ldots & F_{n+1,1}^{(\vec{q}_{n+1})}(y) \\
\vdots & \ddots & \vdots \\
F_{1,n+1}^{(\vec{q}_{1})}(y) & \ldots & F_{n+1,n+1}^{(\vec{q}_{n+1})}(y) \\
\end{array}
\right)
\left(
\begin{array}{c} \fl{x}_1 \\ \vdots \\ \fl{x}_{n+1} \end{array}
\right)
\,.
\ee
Since the multi-degrees $\vec{q}_1,\vec{q}_2,\ldots,\vec{q}_{n+1}$ may be different, these equations are, in general, no longer consistent with the projective scalings on $\mathbb{P}^n[x] \times \vec{\mathbb{P}}[y]$. However, the equations can be understood as defining a complete intersection $\fl{X}$ inside a toric variety,  with a weight system and weights for the defining equations given by
\be
\fl{X} \sim 
\begin{array}{c c c c | c c c c c c}
\fl{x}_1 & \ldots & \fl{x}_{n+1} 	& y 				& \fl{P}_1 & \ldots & \fl{P}_{n+1} & P_{n+2} & \ldots & P_K\\
\hline
1 & \ldots & 1 & 0	 						& 1 & \ldots & 1 & 0 &  \ldots & 0 \\
-\vec{q}_1 & \ldots & -\vec{q}_{n+1} & \square 	& 0 & \ldots & 0 & \vec{q}_{n+2} & \ldots & \vec{q}_K 
\end{array} \,.
\label{eq:111_flp_wgts}
\ee
To the left of the vertical line are the charges of the coordinates under the various scalings, in which $\square$ stands for the unchanged scaling behaviour of the coordinates on $\vec{\mathbb{P}}[y]$, while to the right are the charges of the $K$ polynomials. We note that the entries $-\vec{q}_1 , \ldots, -\vec{q}_{n+1}$ indicate that the projective space $\mathbb{P}^n[\fl{x}]$ is non-trivially fibered over the $\vec{\mathbb{P}}[y]$. 

\medskip

Proving that $\fl{X}$ is indeed the flopped version of $X$ involves two steps: 

\begin{itemize}
\item \textit{Step 1:} Show that there exists a small contraction $\fl{\pi}: \fl{X} \rightarrow X_{\rm sing.}$ with a finite number of rational fibers over the same (singular) points to which $\pi:X\rightarrow X_{\rm sing.}$ contracts its rational fibers.  
\item \textit{Step 2:} Show that the birational map $\phi: X \dashrightarrow \fl{X}$ determined by the composition $(\fl{\pi})^{-1} \circ \pi$ is a flop, in the sense of Definition~\ref{def:flop}, that is there exists a divisor $D$ on $X$ intersecting the contracting curves on $X$ positively, while its image $\phi(D)$ on $\fl{X}$ intersects the contracting curves on $\fl{X}$ negatively. 
\end{itemize}

 Step 1 is easy to argue, as the  analysis carried out above for expressing $X$ as the small resolution of a threefold $\contr{X}\subseteq \vec{\mathbb{P}}[y]$ degenerating at a finite set of points $S$ can also be repeated for $\fl{X}$. Since the discussion relied only on the rank of $F$ and the polynomials $P_{n+2},\ldots, P_K$, which are shared by $X$ and $\fl{X}$, it follows that $\fl{X}$ is also a small resolution of the same singular threefold $\contr{X}$. 

For Step 2,  we construct in Section~\ref{sec:div_img_211} a set of divisors $H_{(i)}$ on $X$ and $\fl{\mathcal{H}}_{(i)}$ on $X'$, with $1\leq i\leq n+1$, such that $\pi(H_{(i)}) = \fl{\pi}(\fl{\mathcal{H}}_{(i)})$ and show in Prop.~\ref{prop:divisors_1} that $H_{(i)}\cdot C = 1$ for any contracting curve $C$ on $X$, while $\fl{\mathcal{H}}_{(i)}\cdot C' = -1$ for any contracting curve $C'$ on $\fl{X}$. These divisors satisfy the final condition in Definition~\eqref{def:flop_def}, and hence the birational map from $X$ to $\fl{X}$ is indeed a flop.

\end{proof}
 
\begin{rmk}
The Calabi-Yau property of $\fl{X}$ follows from that of $X$, since in each row in \eref{eq:111_flp_wgts} the sum of the charges associated with the coordinates equals the sum of the charges associated with the equations. In any given example, one can use techniques of toric geometry to determine the properties of the manifold $\fl{X}$, and in particular to verify that it has the properties expected of the flopped manifold, such as triple intersection numbers and second Chern class related to the originals by \eref{eq:c2_change} (up to a basis transformation), and matching Hodge numbers. The K\"ahler cone of the flopped manifold $\fl{X}$ may, in general, be difficult to obtain. In particular, while the K\"ahler cone of $X$ is known by the assumption that it descends from the ambient space (K\"ahler-favourability), this does not necessarily imply that the same holds for $\fl{X}$.

The description of the flopped manifold given above agrees with that obtained through GLSM reasoning, where the flopped geometry arises as the target space of a second geometric GLSM phase. The flopped manifold $\fl{X}$ is related to the original manifold $X$ by exchanging the weights of the equations with the weights of the coordinates on the embedding space, precisely as one would expect from a GLSM analysis. By contrast, as we will see in \sref{sec:expl_flops2}, the same analysis is not obvious for flops on rows of Type~2.
\end{rmk}

\subsection{Special case: flops to isomorphic manifolds}
\label{sec:special_case}

The equations obtained from the transpose matrix $F^{\mathrm{T}}(y)$ are in general inconsistent with the projective scalings of the original ambient space $\mathbb{P}^n[x] \times \vec{\mathbb{P}}[y]$. However, in the special case when the transpose operation leaves the multi-degrees of the polynomial entries in $F(y)$ unchanged, the new ambient space is the same as the original. 

\begin{prp} The manifolds $X$ and $X'$ are isomorphic (i.e.\ diffeomorphic) if the columns underneath the $1$ entries in \eref{eq:conf_gen_111} are identical, that is, when
\be
\vec{q}_1 = \vec{q}_2 = \ldots = \vec{q}_{n+1} \; .
\ee
\end{prp}

\begin{proof}
The weight system in \eref{eq:111_flp_wgts} is equivalent to that of the original ambient space if the condition is satisfied. In this case, the new defining equations have the same multi-degrees as the original equations and $\fl{X}$ belongs to the same family as $X$. In particular, the coefficients of the defining equations for $\fl{X}$ are simply related to those of $X$ by taking the transpose in \eref{eq:111_mat_eq}. Hence $\fl{X}$ is isomorphic to $X$.
\end{proof}

\begin{ex} The CICY
\be
X_{7761} \sim
\left[
\begin{array}{c | c c c c c }
\mathbb{P}^4 & 1 & 1 & 1 & 1 & 1 \\
\mathbb{P}^4 & 1 & 1 & 1 & 1 & 1 \\
\end{array}
\right] \; .
\label{eq:cicy_7761}
\ee
admits isomorphic flops along both $\mathbb P^4$ directions. 
\end{ex}

We note that, in contrast to the general case, in the isomorphic case the K\"ahler cone of $\fl{X}$ is known (trivially). Its relation to that of $X$ is given by the Picard group isomorphism discussed in \sref{sec:expl_flops_divsandpic}.

\subsection{Divisor images and the Picard group isomorphism}
\label{sec:expl_flops_divsandpic}
The two small contractions $\pi:X\rightarrow X_{\rm sing}$ and $\fl{\pi}:\fl{X}\rightarrow X_{\rm sing}$ discussed above are isomorphisms in codimension one. Denoting by $\{C^{(a)}\}$ and $\{C'^{(a)}\}$ the sets of contracting curves on $X$ and $\fl{X}$ respectively, the map 
\be \phi= (\fl{\pi})^{-1}\circ\pi:X-\{C^{(a)}\} \rightarrow \fl{X}-\{C'^{(a)}\}
\ee
is an isomorphism, hence divisors on $X$ can be uniquely mapped to divisors on $\fl{X}$. In the present section we carry out this analysis for two different classes of divisors that arise naturally. These will be used in the following section to complete the proof that $X$ and $\fl{X}$ are related by a flop.    

We determine the mapping of divisors by matching the images of divisors on the contracted manifold $\contr{X}$ under the two small contractions from $X$ and $\fl{X}$. In the following we require notation for removing rows and columns from the matrix $F(y)$: we write $F_{\hat c_i}(y)$, $F_{\hat r_i}(y)$, and $F_{\hat r_i, \hat c_j}(y)$ respectively for the matrices resulting by removing from $F(y)$ the $i^{\rm th}$ column, the $i^{\rm th}$ row, and simultaneously the $i^{\rm th}$ row and $j^{\rm th}$ column.

\medskip

\begin{defn}
A natural set of divisors $H_{(i)}$ on $X$ is given by intersecting $X$ inside $\mathbb{P}^n[x] \times \vec{\mathbb{P}}[y]$ with the hyperplanes given by the zero loci of the $\mathbb{P}^n[x]$ coordinates, $\{x_i = 0\}$ for $1\leq i\leq n+1$,
\be
H_{(i)} \colon\; \big\{x_i = 0\big\} \cap X~.
\ee
\end{defn}
\medskip

The divisor class of $H_{(i)}$ is $(1,0,\ldots,0)$ in the basis $\{D_j\}$ defined in \sref{sec:CICY}. Since the contracting $\mathbb{P}^1$-curves live inside $\mathbb{P}^n[x]$ at generic locations (which follows from the genericity assumption of Section~\ref{sec:CICY} on the defining polynomials of $X$), the divisors $H_{(i)}$ intersect the $\mathbb{P}^1$s in points. 

The image $\pi(H_{(i)})\subset X_{\rm sing}$ is the locus in $\vec{\mathbb{P}}[y]$ over which $\{x_i = 0\}$ intersects $X$. Looking at the defining equation $F(y) \, \vec{x} = \vec{0}$, since the trivial vector $\vec{x} = \vec{0}$ is not an element in $\mathbb{P}^n[x]$, upon setting $x_i = 0$ a non-trivial solution exists if and only if $\mathrm{rank}\,F_{\hat c_i}(y) < n$. Hence under the small contraction $\pi$ we have the projection
\be
\begin{aligned}
\pi \big( H_{(i)} \big) = 
\big[\!\!
\begin{array}{c | c c c c}
\vec{\mathbb{P}}[y] &
\big( \mathrm{rank} \,F_{\hat c_i}(y) < n \big)
& \vec{q}_{n+2} & \hdots & \vec{q}_K \\
\end{array}
\!\!\big]
\,.
\label{eq:expl_flops_divimg}
\end{aligned}
\ee
Note the condition $\mathrm{rank}\,F_{\hat c_i}(y)< n$ already implies the equation $\mathrm{det}\,F(y) = 0$ which enters in the description of the contracted manifold $\contr{X}$, so the latter equation can be dropped from the description of $\pi ( H_{(i)} )$.

\medskip

Another natural set of divisors on $X$ is as follows. 

\begin{prp} For every $1\leq i\leq n+1$  
\be
\mathcal{H}_{(i)} \colon\; \left\{\frac{\mathrm{det}\,F_{\hat r_i,\hat c_j}(y)}{x_j} = 0 \right\} \cap X~
\ee
is an effective divisor on $X$ containing each of the collapsing curves entirely. 
\end{prp}
\begin{proof}
Since $x_j= 0$ implies that $\mathrm{rank} \,F_{\hat c_j}(y)< n$, it also implies in particular that $\mathrm{det}\, F_{\hat r_i,\hat c_j}(y) = 0$ for any $i$. This means that the zero locus of $x_j$ is contained in that of $\mathrm{det}\, F_{\hat r_i,\hat c_j}(y)$. 

Notably, as we have indicated with the notation of a single index on $\mathcal{H}_{(i)}$, one can check (straightforwardly but slightly tediously) that the locus $\left\{\mathrm{det}\,F_{\hat r_i,\hat c_j}(y) \,/\, x_j = 0 \right\} \cap X$ is independent of the choice of $j$. Moreover, since the locus $\mathrm{det}\, F_{\hat r_i,\hat c_j}(y) = 0$ contains the points over which the collapsing $\mathbb{P}^1$s live (see \eref{eq:111_null2} and observe that ${\rm rank}\,F(y)<n$ implies $F_{\hat r_i,\hat c_j}(y) = 0$ for any $i,j$), this divisor contains each of the collapsing curves entirely, in contrast to the $H_{(i)}$ divisors above which intersected them transversely. 
\end{proof}

To determine the image $\pi(\mathcal{H}_{(i)} )$, note it is straightforward to see that setting $\mathrm{det}\, F_{\hat r_i,\hat c_j}(y) = 0$ and demanding $x_j \neq 0$ in the defining equation $F(y)\,\vec{x} = 0$ implies that $\mathrm{rank}\,F_{\hat r_i}(y)< n$. Hence under the small contraction $\pi$ we have the projection
\be
\begin{aligned}
\pi \big( \mathcal{H}_{(i)} \big) =
\big[\!\!
\begin{array}{c | c c c c}
\vec{\mathbb{P}}[y] &
\big( \mathrm{rank} \,F_{\hat r_i}(y) < n \big)
& \vec{q}_{n+2} & \hdots & \vec{q}_K \\
\end{array}
\!\!\big]
\,.
\label{eq:expl_flops_divimg2}
\end{aligned}
\ee

\medskip

\begin{prp}\label{prop:divisors_1}
There exists a divisor on $X$ intersecting the contracting curves on $X$ positively, while its image under the birational map $\phi:X \dashrightarrow X'$ intersects the contracting curves on $\fl{X}$ negatively. 
\end{prp}
\begin{proof}

On $\fl{X}$ there are the analogous divisors to the above $H_{(i)}$ and $\mathcal{H}_{(i)}$, which we write as $\fl{H}_{(i)}$ and $\fl{\mathcal{H}}_{(i)}$ and which are related by the replacements $x_i \to \fl{x}_i$ and $F(y) \to F(y)^{\mathrm{T}}$. Following the analysis above, their images under the small contraction $\fl{\pi} \colon \fl{X} \to \contr{X}$ are as in \erefs{eq:expl_flops_divimg} and \eqref{eq:expl_flops_divimg2} but with the replacement $F(y) \to F(y)^{\mathrm{T}}$. But since $\mathrm{rank}\,F^{\mathrm{T}}_{\hat c_i}(y) = \mathrm{rank}\,F_{\hat r_i}(y)$, the transpose operation precisely exchanges the two images. Hence we have a matching of divisors on the contracted manifold $\contr{X}$
\be
\pi(H_{(i)}) = \fl{\pi}(\fl{\mathcal{H}}_{(i)})~,\qquad \pi(\mathcal{H}_{(i)}) = \fl{\pi}(\fl{H}_{(i)})~,
\ee
which determines the matching of divisors across the flop
\be
\phi(H_{(i)}) = \fl{\mathcal{H}}_{(i)}~,\qquad \phi(\mathcal{H}_{(i)}) = \fl{H}_{(i)}~.
\ee

Writing divisor classes on $X$ with respect to the basis of the pulled-back hyperplane classes of the projective space factors in the ambient space, and on $\fl{X}$ with respect to the basis of the pullbacks to $\fl{X}$ of the classes corresponding to the rows of the weight system in \eref{eq:111_flp_wgts}, the classes of the above divisors can be read off from their expressions as\footnote{The classes $[H_{(i)}]$ have no dependence on the label $i$ as a consequence of the equal weights of the coordinates $x_i$, i.e.\ this would not be so in the case of a flop in which not only $\fl{X}$ but also $X$ were such that it was embedded in a more general toric ambient space. We also note that the fact that the $[\fl{\mathcal{H}}_{(i)}]$ are exchanged with the $[H_{(i)}]$ under the Picard group isomorphism guarantees that the former too have no dependence on the label $i$.}
\vspace{.2cm}
\be
\begin{array}{ l l l }
X \colon & \big[H_{(i)}\big] = (\,1 \,, \vec{0} \,)
&
\big[\mathcal{H}_{(i)}\big] = \Big( -1 \,, \textstyle\sum_{k = 1}^{n+1} \vec{q}_k - \vec{q}_i \Big) ~,
\vspace{.1cm}\\
X' \colon & \big[\fl{H}_{(i)}\big] = ( \,1 \,, -\vec{q}_i \,\big)
&
\big[\fl{\mathcal{H}}_{(i)}\big] = \Big( -1 \,, \textstyle\sum_{k = 1}^{n+1} \vec{q}_k \Big) ~.
\end{array}
\label{eq:expl_flops_divclmap}
\ee
Since the curve classes of the contracting $\mathbb{P}^1$s on $X$ and $\fl{X}$ are proportional to curve classes dual to the first direction in these two divisor bases, the divisors transversely intersecting the $\mathbb{P}^1$s have positive intersection with the $\mathbb{P}^1$s while the divisors containing the $\mathbb{P}^1$s have negative intersection, as expected.
\end{proof}

\begin{crl}\label{crl:Pic_iso}
The Picard group isomorphism between ${\rm Pic}(X)$ and ${\rm Pic}(\fl{X})$ is the map that exchanges the divisor classes as ${[H_{(i)}]\leftrightarrow[\fl{\mathcal{H}}_{(i)}]}$ and ${[\mathcal{H}_{(i)}]\leftrightarrow[\fl{H}_{(i)}]}$ and which trivially maps classes with no component in the first entry, namely
\be
\mathrm{Pic}(X) \to \mathrm{Pic}(\fl{X}) \, \colon
\hspace{.2cm}
\vec{v} \, \mapsto
\left( \begin{array}{c c}
-1 & \vec{0}^{\,\mathrm{T}} \\[4pt]
\sum_{k = 1}^{n+1} \vec{q}_k  & \mathbbm{1}
\end{array} \right)
\vec{v} \,.
\label{eq:expl_flops_piciso}
\ee
\end{crl}

\subsection{Counting the number of contracting curves}
\label{sec:expl_flops_count}

There are at least two ways to count the number of $\mathbb{P}^1$s involved in the small resolution. One method is to compare the Euler characteristic of the original manifold $X$ with that of the smoothed contracted manifold $\defm{X}$. Since the Euler characteristic adds in the surgery that replaces the nodal points with $\mathbb{P}^1$s, and $\ec(\mathbb{P}^1)=2$, one has
\be
\mathrm{num}\big(\mathbb{P}^1\big) = \tfrac{1}{2}\big(\ec(X) - \ec(\defm{X})\big) \,.
\ee
These Euler characteristics are straightforward to compute from the configuration matrices of $X$ and $\defm{X}$ in \erefs{eq:conf_gen_111} and \eqref{eq:111_null11}, the latter of which follows immediately from the former.

A second method is to use the Giambelli-Thom-Porteous formula, which in our context reduces to the following. Define the line bundle sum $F = \bigoplus_{j=1}^{n+1} \mathcal{O}_{\vec{\mathbb{P}}}(\vec{q}_j)$ and consider its restriction $\mathcal{F}$ to $[\,\vec{\mathbb{P}}[y] ~|~ \vec{q}_{n+2} ~ \hdots ~ \vec{q}_K ]$. Writing $c_i(\mathcal{F})$ for the (Poincare dual of the) $i^{\rm th}$ Chern class of $\mathcal{F}$, the number of singular points of the variety in \eref{eq:111_null1}, or equivalently the number of $\mathbb{P}^1$s in the small resolution, is
\be
\mathrm{num}\big(\mathbb{P}^1\big) = 
\left|
\begin{array}{c c}
c_2(\mathcal{F}) & c_3(\mathcal{F}) \\
c_1(\mathcal{F}) & c_2(\mathcal{F})  
\end{array}
\right| 
=
\left|
\begin{array}{c c}
c_2(F) & c_3(F) \\
c_1(F) & c_2(F)  
\end{array}
\right| 
\,\cdot
\prod_{j=n+2}^{K} \Big( \, \vec{q}_j \cdot \vec{H} \, \Big) \,.
\ee
In the last expression all intersections are taken on $\vec{\mathbb{P}}[y]$, and the product factor is a series of intersections which implement the restriction to $[\,\vec{\mathbb{P}}[y] ~|~ \vec{q}_{n+2} ~ \hdots ~ \vec{q}_K ]$, in which $\vec{H}$ is a list of the hyperplane classes of each of the projective spaces in $\vec{\mathbb{P}}[y]$. 

\begin{ex} Consider
\be
X_{7807} \sim
\left[
\begin{array}{c | c c c c }
\mathbb{P}^2 & 1 & 1 &1 & 0 \\
\mathbb{P}^5 & 2 & 1 & 1 & 2 \\
\end{array}
\right]
\,.
\ee
so that $F = \mathcal{O}_{\mathbb{P}^5}(2) \oplus \mathcal{O}_{\mathbb{P}^5}(1) \oplus \mathcal{O}_{\mathbb{P}^5}(1)$ and $\prod_{j=n+2}^{K} \big(\vec{q}_j \cdot \vec{H}\big) = 2H$. The Chern classes of $F$ are then
\be
c_1(F) = 4H \,, \quad c_2(F) = 5H^2 \,, \quad c_3(F) = 2H^3 \,.
\ee
and the number of collapsing curves is 
\be
\mathrm{num}(\mathbb{P}^1) =
\left|
\begin{array}{c c}
5H^2 & 2H^3 \\
4H & 5H^2
\end{array}
\right|  \cdot (2H) = 34 \,.
\ee
\end{ex}

\begin{rmk}
In the special case discussed in \sref{sec:special_case} when the flopped Calabi-Yau $\fl{X}$ is isomorphic to the original Calabi-Yau~$X$, one can also compute the number of collapsing $\mathbb{P}^1$s by leveraging this isomorphism in the relation between the second Chern class of the original and flopped manifolds in \eref{eq:c2_change}. In particular, since we know the second Chern class after the transition, we can use this relation to learn about the contracting $\mathbb{P}^1$ curves. Taking the intersection on the right of \eref{eq:c2_change} with ${H_{(i)} \sim (1,0)}$ and on the left with ${\fl{\mathcal{H}}_{(i)} \sim \big( -1 \,,\, \vec{q}_1 + 2\sum_{k=2}^n \vec{q}_k \big)}$, i.e.\ using the Picard group isomorphism in \eref{eq:expl_flops_piciso}, we have
\be
\begin{aligned}
2 \, \sum_a C^{(a)} \cdot H_{(i)} &= c_2(X) \cdot \fl{\mathcal{H}}_{(i)} - c_2(X)\cdot H_{(i)} \\
\Rightarrow ~~ \mathrm{num}\big(\mathbb{P}^1\big) &= \tfrac{1}{2} \, c_2(X) \cdot \big( \fl{\mathcal{H}}_{(i)} - H_{(i)} \big) \\
&= \tfrac{1}{2} \, c_2(X) \cdot \big( -2 \,,\, \vec{q}_1 + 2\sum_{k=2}^n \vec{q}_k \big) \,.
\end{aligned}
\ee
\end{rmk}

 \begin{ex} Consider the CICY in \eref{eq:cicy_7761}, for which one can check that $c_2(X) = (50,50)$, one has $\fl{\mathcal{H}}_{(i)} \sim (-1,4)$, so that one finds $\mathrm{num}\big(\mathbb{P}^1\big) = 50$.
\end{ex}

\section{Performing flops on rows of Type 2}
\label{sec:expl_flops2}
A configuration matrix containing a row of Type~2 in the classification of \sref{sec:class} is, up to permutations of rows and columns, of the form
\be
X \sim 
\left[
\begin{array}{c | c c c c c c c}
\mathbb{P}^n[x] & 2 & 1 & \ldots & 1 & 0 & \ldots & 0 \\
\vec{\mathbb{P}}[y] & \vec{q}_1 & \vec{q}_2 & \ldots & \vec{q}_{n} & \vec{q}_{n+1} & \hdots & \vec{q}_K \\
\end{array}
\right]
\,.
\label{eq:conf_gen_211}
\ee
The coordinates on the first projective space are denoted by $x=[x_1:x_2:\ldots:x_{n+1}]$, while $\vec{\mathbb{P}}[y]$ represents the product of the remaining projective spaces with coordinates collectively denoted by $y$. The ambient space is then of the form $\mathcal{A}=\mathbb{P}^n\times\vec{\mathbb{P}}$. The vector $\vec{q}_i$ collects the entries in the $i^{\rm th}$ column of the configuration matrix after dropping the first row. 

\subsection{The contraction maps}
We begin the discussion of flops on rows of Type 2 by studying the small contraction maps that can arise in this case. 

\begin{prp}\label{prp:contractions}
A generic K\"ahler-favourable CICY threefold whose configuration matrix contains a row of Type 2 admits two small contractions $\pi^-,\pi^+:X\rightarrow X_{\rm sing}$, where $X_{\rm sing}$ has isolated singularities. The composition $(\pi^-)^{-1}\circ \pi^+$ induces a birational map between $X$ and itself.
\end{prp}

\begin{proof}
The last $K-n$ equations in the configuration matrix in \eref{eq:conf_gen_211} define a complete intersection $Y$ inside $\vec{\mathbb{P}}[y]$,
\be
Y \sim 
\left[
\begin{array}{c | c c c }
\vec{\mathbb{P}}[y] & \vec{q}_{n+1} & \hdots & \vec{q}_K \\
\end{array}
\right] \;,
\ee
(unless $K=n+1$, in which case simply $Y = \vec{\mathbb{P}}[y]$) which one can check is a threefold. The first $n$ equations in the configuration matrix have degrees $\{2,\,1,\,\ldots,\,1\}$ in the $x$-coordinates, and hence over a generic point in $Y$ they determine two points inside $\mathbb{P}^n[x]$. Hence the CICY is almost everywhere a double cover of the manifold $Y$. Over a codimension-one locus inside $Y$, the two points inside $\mathbb{P}^n[x]$ coincide. Further, over a codimension-three locus inside $Y$, i.e.\ at a set of points, the $n$ equations become degenerate, admitting as solutions an entire~$\mathbb{P}^1$ inside $\mathbb{P}^n[x]$. Hence, $X$ is the small resolution of a branched double cover of $Y$ at a number of singular points.

\begin{figure}[H]
\begin{center}
\begin{tikzpicture}[
    scale=1.5,
    axis/.style={very thick, ->, >=stealth'},
    important line/.style={thick},
    dashed line/.style={dashed, thick},
    every node/.style={color=black,}
 ]
\draw[->] (0,0)  -- (6,0) node(xline)[]{};
\draw[->] (0,0) -- (0,2.5) node(yline)[]{};
\draw[domain=.4:2,black,smooth,very thick] plot ({\x},{2.6-1.2*\x+.3*\x^2});
\draw[domain=.4:2,black,smooth,very thick] plot ({\x},{0.2+1.2*\x-.3*\x^2});
\draw[domain=2:4,black,smooth,very thick] plot ({\x},{1.4});
\draw[domain=4:6,black,smooth,very thick] plot ({\x},{1.4+(\x-4)^2-0.4*(\x-4)^3});
\draw[domain=4:6,black,smooth,very thick] plot ({\x},{1.4-(\x-4)^2+0.4*(\x-4)^3});
\draw[black,very thick] (3,.4) to (3,2.4);
\fill[black] (3,0) circle (1pt) node at (3.25,.2) {$y^*$};
\node at (6.35,0) {$\vec{\mathbb{P}}[y]$};
\node at (-.4,2.35) {$\mathbb{P}^n[x]$};
\node at (3.25,2.4) {$\mathbb{P}^1$};
\end{tikzpicture}
\end{center}
\caption{Any (generic, K\"ahler-favourable) complete-intersection Calabi-Yau threefold whose configuration matrix is as in \eref{eq:conf_gen_211} is a small resolution of a branched double cover of $\vec{\mathbb{P}}[y]$.}
\end{figure}
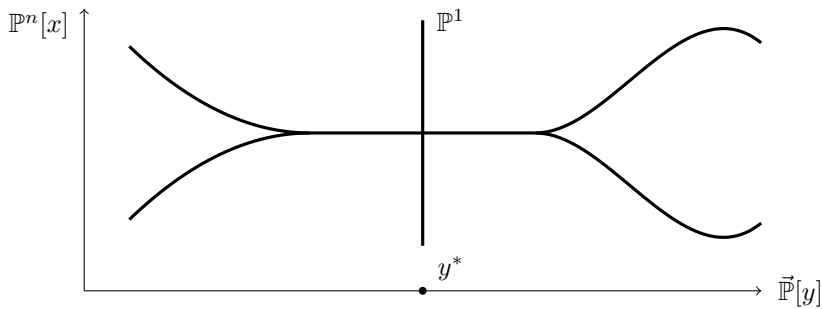

To describe this explicitly, we write out the first $n$ equations defining $X$ in the form
\be
\begin{gathered}
0 = \vec{x}^{\,\mathrm{T}} A(y) \vec{x} =
\begin{pmatrix} \, x_1 & \ldots & x_{n+1} \,\end{pmatrix}
\left(\begin{array}{c c c}
A_{1,1}^{(\vec{q}_1)}(y) & \ldots & A_{1,n+1}^{(\vec{q}_1)}(y) \\
\vdots & \ddots & \vdots \\
A_{n+1,1}^{(\vec{q}_1)}(y) & \ldots & A_{n+1,n+1}^{(\vec{q}_1)}(y) \\
\end{array}\right)
\begin{pmatrix} x_1 \\ \vdots \\ x_{n+1} \end{pmatrix} \,,
\vspace{.2cm}\\
\vec{0} = B(y) \vec{x} =
\left(\begin{array}{c c c}
B_{1,1}^{(\vec{q}_2)}(y) & \ldots & B_{1,n+1}^{(\vec{q}_2)}(y) \\
\vdots & \ddots & \vdots \\
B_{n-1,1}^{(\vec{q}_n)}(y) & \ldots & B_{n-1,n+1}^{(\vec{q}_n)}(y) \\
\end{array}\right)
\begin{pmatrix} x_1 \\ \vdots \\ x_{n+1} \end{pmatrix} 
=
\begin{pmatrix} \vec{B}_1^{\mathrm{T}} \\ \vdots \\ \vec{B}_{n-1}^{\mathrm{T}} \end{pmatrix}
\begin{pmatrix} x_1 \\ \vdots \\ x_{n+1} \end{pmatrix} \,,
\end{gathered}
\ee

\noindent
where the superscripts denote again the multi-degrees of the functions in $y$ and the last equality should be read as a definition of the vectors $\vec{B}_i$. Note that the matrix $A$ is square (and symmetric) while $B$ is an $(n-1) \times (n+1)$ matrix so that the second line describes $n-1$ equations.

Over a generic point $y$ these equations pick out two points inside $\mathbb{P}^n[x]$ and these solutions coincide when the discriminant $\Delta(y)$ vanishes. For $n>1$ one can check that this discriminant is given by
\be
\Delta(y) = \sum_{\sigma,\rho \,\in S_{n+1}}C_2(A)_{(\sigma(1),\ldots,\sigma(n-1)),(\rho(1),\ldots,\rho(n-1))} B_{1,\sigma(1)}B_{1,\rho(1)} \ldots B_{n-1,\sigma(n-1)}B_{n-1,\rho(n-1)} \,.
\ee
Here $C_2(A)_{(\sigma(1),\ldots,\sigma(n-1)),(\rho(1),\ldots,\rho(n-1))}$ is the cofactor of $A$ corresponding to the $2 \times 2$ submatrix which results upon removing rows $\sigma(1),\ldots,\sigma(n-1)$ and columns $\rho(1),\ldots,\rho(n-1)$. The multi-degree of $\Delta$ is $2\sum_{i=1}^n \vec{q}_i$. When $n=1$, we only have the quadratic equation and the discriminant is given by $\Delta(y) = \mathrm{det}\,A(y)$.

Further, for $n>1$, there are points within the branch locus at which the defining equations become degenerate, allowing a one-dimensional space of solutions. This can happen in two ways.
\begin{itemize}
\item[(1)] The quadratic equation factors at a point $y^*\in \vec{\mathbb{P}}[y]$ into two linears, one of which is linearly dependent with the $n-1$ linear equations defined by $B$, that is
\be
\vec{x}^{\,\mathrm{T}}  A(y^*)  \vec{x} = ( \ldots ) \, \vec{c}^{\,\mathrm{T}}  \vec{x}
\qquad\mathrm{ and }\qquad
\mathrm{rank}
\big(\!\!\!\begin{array}{c c c c} \vec{B}_1(y^*) & \ldots & \vec{B}_{n-1}(y^*) & \!\!\vec{c} \end{array} \!\!\big)
\leq n - 1
 \,,
 \label{eq:1c_curves}
\ee
for some $\vec{c} = ( \, c_1 \,,\, \ldots \,,\, c_{n+1} \,)^{\mathrm{T}} \in \mathbb{C}^{n+1}$. The equations then determine an entire $\mathbb{P}^1$ inside $\mathbb{P}^n[x]$. Being defined by linear equations, this rational curve intersects a generic hyperplane in  $\mathbb{P}^n[x]$ in precisely one point. We denote by $[C]$ its curve class in $X$.

\item[(2)] The $n-1$ linear equations become linearly dependent at a point $y^*\in \vec{\mathbb{P}}[y]$, that is
\be
\quad
\mathrm{rank }\, B(y^*) \leq n-2 \,.
\label{eq:2c_curves}
\ee
In this case the equations defining the $\mathbb{P}^1$ inside $\mathbb{P}^n[x]$ include a quadratic, and hence the corresponding curve is in the class $2[C]$.

\end{itemize}

\noindent For the special case $n=1$, in which $B$ is absent, the contracting $\mathbb{P}^1$s occur where the three independent components of the symmetric $2\times 2$ matrix $A$ vanish, and these $\mathbb{P}^1$s are all in the same curve class $[C]$.

\medskip

Contracting the $\mathbb{P}^1$s leaves a branched double cover of $Y$ with branch locus over $\Delta(y)=0$. To explicitly describe this double cover of $Y$ we add an extra coordinate $\xi$ to the ambient space $\vec{\mathbb{P}}[y]$ of $Y$, and introduce the equation
\be
\xi^2 - \Delta(y) = 0 \,.
\label{eq:sing_211}
\ee
The new ambient space is a toric variety for which the weight of $\xi$ is chosen to make the above equation consistent. Altogether, the singular manifold resulting from the small contraction of $X$ is then described as a complete intersection in a toric ambient space, namely
\be
\contr{X} \sim
\begin{array}{c c | c c c c}
\xi  						& y			& & & & \\
\hline \rule{0pt}{2.5ex}
\sum_{i=1}^n \vec{q}_i	& \square 	& \left( \xi^2 - \Delta(y) = 0 \right) & \vec{q}_{n+1} & \hdots & \vec{q}_K  \\
\end{array}
\ee
where $\square$ stands for the unchanged scaling behaviour of the coordinates on the $\vec{\mathbb{P}}[y]$, and where the $K-n$ equations with multi-degrees $\vec{q}_{n+1}$ to $\vec{q}_K$ are the same as those appearing in the last $K-n$ equations in \eqref{eq:conf_gen_211}.

Since the hypersurface in \eref{eq:sing_211} is invariant under $\xi \to - \xi$, there are actually two contraction maps from $X$ to $\contr{X}$,
\begin{equation}\label{eq:comm_diag_XX}
\begin{aligned}
\begin{tikzpicture}
\node (X) at (0,0) {$X$};
\node (Xt) at (3,0) {$X$};
\node (Xs) at (1.5,-1.3) {$\;X_{\rm sing}$};
\node (pim) at (.4,-.7) {$\pi^-$};
\node (pip) at (2.6,-.7) {$\pi^+$};
\draw[<->,dashed] (X)+(.4,0) -- (Xt) 
	node[midway, above] {$\;\phi$};
\draw[->,] (X)+(.3,-0.2) -- (Xs);
\draw[->,] (Xt) -- (Xs); 
\end{tikzpicture}
\end{aligned}
\end{equation}
where the projections $\pi^-$ and $\pi^+$ agree on the preimage of the branch locus, and outside of this they agree upon composition with the automorphism $\xi\rightarrow -\xi$ on $X_{\rm sing}$. The composition $(\pi^-)^{-1} \circ \pi^+$ induces a birational map $\phi$ from $X$ to itself. 
\end{proof}

Below we show that $\phi$ is in fact a flop, by first determining in \sref{sec:div_img_211} the mapping of divisors under $\phi$, and so constructing a divisor $D$ on $X$ such that the collapsing curves have a positive intersection with $D$ and a negative intersection with the corresponding divisor $D'$ under $\phi$.

\begin{rmk}
One can check that deformations of \eref{eq:sing_211} to include a linear term in $\xi$ give rise to smooth hypersurfaces, and hence that $\contr{X}$ belongs to a family $\defm{X}$ of smooth Calabi-Yau threefolds, which are complete intersections in a toric ambient space, namely
\be
\defm{X} \sim
\begin{array}{c c | c c c c}
\xi  						& y			& & & & \\
\hline \rule{0pt}{2.5ex}
\sum_{i=1}^n \vec{q}_i	& \square 	& 2\sum_{i=1}^n \vec{q}_i & \vec{q}_{n+1} & \hdots & \vec{q}_K  \\
\end{array}
\ee
Hence the threefold $X$ in \eref{eq:conf_gen_211} is on the resolution branch of a conifold transition
\be
\begin{array}{c c c c c}
\defm{X} & \longleftrightarrow & \contr{X} & \longleftrightarrow & X
\end{array}
\ee
and when we construct the flopped space below, we will be constructing the other possible resolution branch of this conifold transition.
\end{rmk}

\subsection{Divisor images and the Picard group isomorphism}
\label{sec:div_img_211}
We begin with the following definition.

\begin{defn}
A natural set of divisors $H_{(i)}$ on $X$ is given by intersecting $X$ inside $\mathbb{P}^n[x] \times \vec{\mathbb{P}}[y]$ with the hyperplanes given by the zero loci of the $\mathbb{P}^n[x]$ coordinates, $\{x_i = 0\}$ for $1\leq i\leq n+1$,
\be
H_{(i)} \colon\; \big\{x_i = 0\big\} \cap X~.
\ee
\end{defn}

\begin{prp}\label{prop:divisors_2}
There exists a divisor on $X$ intersecting the contracting curves positively, while its image under $\phi$ intersects the contracting curves negatively. 
\end{prp}
\begin{proof}
The divisor class of $H_{(i)}$ is $(1,0,\ldots,0)$ in the basis $\{D_i\}$ defined in \sref{sec:CICY}. This divisor $H_{(i)}$ intersects the contracting $\mathbb{P}^1$s in points, and in particular, it intersects the $\mathbb{P}^1$s of class $C$ and $2C$, described in \erefs{eq:1c_curves} and \eqref{eq:2c_curves}, in one and two points respectively. 

The image $\pi^+(H_{(i)})\subset X_{\rm sing}$ is the locus in $\vec{\mathbb{P}}[y]$ over which $\{x_i = 0\}$ intersects $X$. Note that if $x_i = 0$ then from the equation $B(y) \, \vec{x} = 0$ one has either $\vec{x} \propto \hat{e}_i \wedge \vec{B}_1 \wedge \ldots \wedge \vec{B}_{n-1}$ or $\mathrm{det}\,\left(\hat{e}_i \,, \vec{B}_1 \,, \ldots \,, \vec{B}_{n-1}\right) = 0$, where $\hat{e}_i$ is the $i^\mathrm{th}$ unit vector. Combining this with the equation $\vec{x}^{\,\mathrm{T}} A \, \vec{x} = 0$, it follows that
\be
\begin{gathered}
\pi^+(H_{(i)}) = 
\big[\!\!
\begin{array}{c | c c c c}
\vec{\mathbb{P}}[y] &
(f_{(i)}(y) =0)
& \vec{q}_{n+2} & \hdots & \vec{q}_K \\
\end{array}
\!\!\big]
\,, \\
\mathrm{where} \hspace{.2cm} f_{(i)}(y) = \left( \hat{e}_i  \wedge \vec{B}_1(y) \wedge \ldots \wedge \vec{B}_{n-1}(y) \right)^{\mathrm{T}}  A(y) \left( \hat{e}_i \wedge \vec{B}_1(y) \wedge \ldots \wedge \vec{B}_{n-1}(y) \right) \,.
\end{gathered}
\ee
However over $f_{(i)}(y) = 0$ there are two points on the Calabi-Yau (at least away from the branch locus). Hence by taking the locus in $X$ over $f_{(i)}(y) = 0$ but removing the points $x_i=0$, one constructs a second effective divisor, described by 
\be
\mathcal{H}_{(i)} \colon\; \left\{ \frac{f_{(i)}(y)}{x_i} = 0 \right\}\cap X~.
\ee
Away from the branch locus $H_{(i)}$ and $\mathcal{H}_{(i)}$ pick out distinct points. Within the branch locus but away from the $\mathbb{P}^1$s the two loci coincide. However there is a significant qualitative difference at the $\mathbb{P}^1$ loci, since while $H_{(i)}$ intersects the contracting $\mathbb{P}^1$s transversely, $\mathcal{H}_{(i)}$ contains these curves entirely.

In the projection to $\contr{X}$, the $\mathbb{P}^1$s are contracted to points and the qualitative difference between $H_{(i)}$ and $\mathcal{H}_{(i)}$ disappears: these divisors simply map to opposite branches of the singular double cover, i.e.\ the images $\pi^+(H_{(i)})$ and $\pi^+(\mathcal{H}_{(i)})$ are related by the $\mathbb{Z}_2$ symmetry $\xi \to - \xi$. However, the second contraction $\pi^-$ is related to the first $\pi^+$ by composition with this symmetry. Hence
\be
\pi^+ (H_{(i)}) = \pi^- (\mathcal{H}_{(i)})~,\qquad \pi^+ (\mathcal{H}_{(i)}) = \pi^- (H_{(i)})~,
\ee
and, consequently, under the induced birational map $\phi \colon X \dashrightarrow X$, we have the divisor mapping
\be
\phi(H_{(i)}) = \mathcal{H}_{(i)}~,\qquad \phi(\mathcal{H}_{(i)}) = H_{(i)}~.
\ee

Writing divisor classes on $X$ with respect to the basis of the pulled-back hyperplane classes of the projective space factors in the ambient space, the classes of the above divisors can be read off from their expressions as
\vspace{.2cm}
\be
\big[H_{(i)}\big] = (\,1 \,, \vec{0} \,)
~, \quad
\big[\mathcal{H}_{(i)}\big] = \Big(-1 \,,\,\vec{q}_1 + 2\textstyle\sum_{k=2}^n \vec{q}_k \Big) ~.
\ee
The divisors transversely intersecting the $\mathbb{P}^1$s have positive intersection with the $\mathbb{P}^1$s while the divisors containing the $\mathbb{P}^1$s have negative intersection, as expected. 
\end{proof}

\begin{crl}\label{crl:picard_iso2}
The Picard group isomorphism between $X$ and itself is the map that exchanges the divisor classes as ${[H_{(i)}]\leftrightarrow[\mathcal{H}_{(i)}]}$ and which trivially maps classes with no component in the first entry, namely
\be
\mathrm{Pic}(X) \to \mathrm{Pic}(X) \, \colon
\hspace{.2cm}
\vec{v} \, \mapsto
\left( \begin{array}{c c}
-1 & \vec{0}^{\,\mathrm{T}} \\[4pt]
\vec{q}_1 + 2\sum_{k=2}^n \vec{q}_k & \mathbbm{1}
\end{array} \right)
\vec{v} \,.
\label{eq:piciso_2}
\ee
\end{crl}

\subsection{The flop and the new K\"ahler cone}
\label{sec:flop_piciso_211}
We are now in a position to prove the following.

\begin{thm}\label{thm:flops2}
A generic CICY threefold with a row of Type~2 admits a flop to a threefold isomorphic (i.e.\ diffeomorphic) to itself. Moreover, this is in fact an isomorphism of complex manifolds.
\end{thm}
\begin{proof}
Under the birational map from $X$ to itself, which contracts and then resolves the $\mathbb{P}^1$s and gives a manifold isomorphic to the original Calabi-Yau, the divisor $H_{(i)}$ maps to the divisor $\mathcal{H}_{(i)}$ (and vice versa). Moreover, the intersections of these divisors with the contracting $\mathbb{P}^1$s are positive and negative, respectively. Hence by Definition~\eqref{def:flop_def}, the associated birational map from $X$ to itself is a flop.
\end{proof}

We see that flops on rows of Type~2 always produce an isomorphic Calabi-Yau threefold. This is in contrast to the case of flops on rows of Type~1 treated in \sref{sec:expl_flops}, where the flopped Calabi-Yau is generically topologically distinct from the original one (except in the special case noted in \sref{sec:special_case}).

As in the special case of isomorphic flops from Type~1 rows, since all flops on rows of Type~2 produce an isomorphic Calabi-Yau the K\"ahler cone of the flopped manifold is known (trivially), and its relation to that of the original Calabi-Yau is given by the Picard group isomorphism in \eref{eq:piciso_2}.

\begin{ex}
As a first example, consider a generic Calabi-Yau threefold $X$ with configuration matrix and defining equation given by\footnote{We note that the birational geometry of hypersurfaces in products of projective spaces has also been described in Ref.~\cite{Ottem2013BirationalGO}. This analysis overlaps with the cases treated in the present article in three cases: the present example, the hypersurface of multi-degree $(2,2,3)$ in $\mathbb{P}^1 \times \mathbb{P}^1 \times \mathbb{P}^2$, and the hypersurface of multi-degree $(2,2,2,2)$ in $\mathbb{P}^1 \times \mathbb{P}^1 \times \mathbb{P}^1 \times \mathbb{P}^1$.}
\be
X=X_{7887}
~\sim\,
\left[
\begin{array}{c | c }
\mathbb{P}^1[x] & 2 \\
\mathbb{P}^3[y] & 4 \\
\end{array}
\right]
\,:~~\qquad
\left\{ \begin{array}{c} P^{(4)}(y)\,x_1^2 + 2Q^{(4)}(y)\,x_1x_2 + R^{(4)}(y)\,x_2^2 = 0 \end{array} \right\} \,. \\
\label{eq:cicy_7887}
\ee

For a generic $y\in\mathbb{P}^3$ the defining equation gives two solutions for $x$, while when the discriminant $P(y)R(y)-Q(y)^2$ vanishes, the two solutions coincide. On the other hand, when $P(y)=Q(y)=R(y)=0$, which happens at $64$ points in $\mathbb{P}^3$, the equation admits an entire $\mathbb{P}^1$ as a solution. By contracting these curves one obtains the singular double cover, which can be described as a hypersurface in a weighted projective space:
\be
X_{\rm sing} = \{(\xi,y_1,y_2,y_3,y_4)\in \mathbb{P}_{[4:1:1:1:1]}~|~ \xi^2-(PR-Q^2)(y)=0\}~.
\ee
The singularities of $\contr{X}$ are of $A_1$ type, so locally these correspond to the Atiyah flop.

The Picard group isomorphism, written in the basis $\{[D_1],[D_2]\}$ of the pulled-back hyperplane classes on $\mathbb{P}^1$ and $\mathbb{P}^2$, respectively, is given by the matrix
\be
\mathrm{Pic}(X) \to \mathrm{Pic}(X) \, \colon
\hspace{.2cm}
\vec{v} \, \mapsto
\left( \begin{array}{c c}
\!\!-1 & 0 \\ ~4 &1
\end{array}\right)
\vec{v} \,.
\ee
In particular, this means that $\phi([D_1])=-[D_1]+4[D_2]$ and $\phi([D_2])=D_2$. The map can be extended by linearity to ${\rm Pic}(X)\otimes \mathbb R$ giving the following picture which involves two cones: the K\"ahler cone ${\mathcal{K}(X) = \langle (1,0) \,,\, (0,1) \rangle}$ and its image ${\phi(\mathcal{K}(X)) = \langle (-1,4) \,,\, (0,1) \rangle}$, forming together the extended K\"ahler cone.
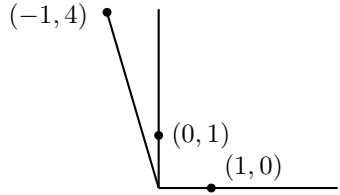
\begin{center}
\begin{tikzpicture}[
    scale=1.4,
    axis/.style={very thick, ->, >=stealth'},
    important line/.style={thick},
    dashed line/.style={dashed, thick},
    every node/.style={color=black,}
 ]
\draw[-, thick] (0,0)  -- (1.7,0) node(xline)[]{};
\draw[-,thick] (0,0) -- (0,1.7) node(yline)[]{};
\draw[-, thick] (0,0) -- (-0.5,1.7) node(yline)[]{};
\fill[black] (.5,0) circle (1.2pt) node at (.9,.2) {$(1,0)$};
\fill[black] (0,.5) circle (1.2pt) node at (.4,.5) {$(0,1)$};
\fill[black] (-0.49,1.67) circle (1.2pt) node at (-1.05,1.63) {$(-1,4)$};
\end{tikzpicture}
\end{center}
\end{ex}

\begin{ex}
For a second example, we choose a generic Calabi-Yau threefold $X$ with configuration matrix and defining equation given by
\begin{equation*}
X=X_{7883}
~\sim\,
\left[
\begin{array}{c | c  c}
\mathbb{P}^2[x] & 2 &1\\
\mathbb{P}^3[y] & 3 & 1 \\
\end{array}
\right]
\,:~\qquad
\left\{
\begin{array}{c}
\vec{x}^{\,\mathrm{T}} A \,\vec{x} = 0 \\
\vec{B}^{\mathrm{T}} \vec{x} = 0
\end{array}
 \right\} \,,
 \label{eq:cicy_7883}
\end{equation*}
where $A$ is a symmetric $3 \times 3$ matrix whose entries are degree three polynomials in $y$, while $\vec{B}$ is a 3-dimensional vector whose entries are degree one polynomials in $y$.

The second equation can be used to eliminate one of the $x$-variables from the first equation. The resulting quadratic equation has a determinant which can be written as
\be
\Delta(y) = \vec{B}(y)^{\mathrm{T}} C_A (y) \vec{B}(y)~,
\ee
where $C_A$ is the matrix of cofactors of $A$. The singular double cover at the intermediate point of the flop is again a hypersurface in a weighted projective space:
\be
X_{\rm sing} = \{(\xi,y_1,y_2,y_3,y_4)\in \mathbb{P}_{[4:1:1:1:1]}~|~ \xi^2-\Delta(y)=0\}~.
\ee
The threefold $X_{\rm sing}$ has $73$ singular points and the map $X\rightarrow X_{\rm sing}$ contracts $72$ curves in the class $[C_1]$ and $1$ curve in the class $2[C_1]$, where $[C_1],[C_2]$ are the curve classes dual to the hyperplane classes $[D_1]$, $[D_2]$ of $\mathbb{P}^2$ and, respectively, $\mathbb{P}^3$ (pulled-back to the Calabi-Yau $X$).

The Picard group isomorphism written in the basis $\{[D_1],[D_2]\}$ is given by the matrix
\be
\mathrm{Pic}(X) \to \mathrm{Pic}(X) \, \colon
\hspace{.2cm}
\vec{v} \, \mapsto
\left( \begin{array}{c c}
\!\!-1 & 0 \\ ~5 &1
\end{array}\right)
\vec{v} \,,
\ee
giving an extended K\"ahler cone qualitatively as in the previous example but with a boundary ray $(-1,5)$. Note that the singular manifold involved in this flop belongs to the same deformation class as that in the previous example, hence the threefolds $X_{7887}$ and $X_{7883}$ are connected by two conifold transitions. 
\end{ex}

\subsection{Counting the number of contracting curves}
\label{sec:counting_curves_type2}
Unlike in the case of flops on rows of Type~1 treated in \sref{sec:expl_flops}, in the case of flops on rows of Type~2 there are two types of $\mathbb{P}^1$s involved in the contraction: those with class $[C]$ and those with class $2[C]$. 

\begin{prp}
The numbers $\mathrm{num}\big(\mathbb{P}^1_{[C]}\big)$ and $\mathrm{num}\big(\mathbb{P}^1_{2[C]}\big)$ of contracting curves in each class are given by the two relations
\be
\begin{gathered}
\mathrm{num}\big(\mathbb{P}^1_{[C]}\big) + \mathrm{num}\big(\mathbb{P}^1_{2[C]}\big) = \tfrac{1}{2}\big(\ec(X) - \ec(\defm{X})\big)
\,, \\
\mathrm{num}\big(\mathbb{P}^1_{[C]}\big) + 2 \,\mathrm{num}\big(\mathbb{P}^1_{2[C]} \big) = \tfrac{1}{2} \, c_2(X) \cdot \Big( -2 \,,\, \vec{q}_1 + 2\sum_{k=2}^n \vec{q}_k \Big) \,.
\end{gathered}
\ee
\end{prp}

\begin{proof}
The sum of the total numbers of both types can be extracted from the difference in Euler characteristic between $X$ and the smoothed contracted manifold $\defm{X}$,
\be
\mathrm{num}\big(\mathbb{P}^1_{[C]}\big) + \mathrm{num}\big(\mathbb{P}^1_{2[C]} \big) = \tfrac{1}{2}\big(\ec(X) - \ec(\defm{X})\big) \,.
\label{eq:expl_flops2_euldif}
\ee
One may also want to know the two numbers $\mathrm{num}\big(\mathbb{P}^1_C\big)$ and $\mathrm{num}\big(\mathbb{P}^1_{2C}\big)$ independently. One way to compute this is by leveraging the fact that the flopped manifold is isomorphic to the original in the relation for the difference in second Chern classes in \eref{eq:c2_change}. In particular, we immediately know the second Chern class after the transition, and hence we can use this relation to learn about the contracting $\mathbb{P}^1$ curves. Taking the intersection on the right of \eref{eq:c2_change} with $H_{(i)} \sim (1,0)$ and on the left with $\mathcal{H}_{(i)} \sim \big( -1 \,,\, \vec{q}_1 + 2\sum_{k=2}^n \vec{q}_k \big)$, i.e.\ using the Picard group isomorphism in \eref{eq:piciso_2}, we have
\be
\begin{aligned}
2 \, \sum_a C^{(a)} \cdot H_{(i)} &= c_2(X) \cdot \mathcal{H}_{(i)} - c_2(X) \cdot H_{(i)}
\\[.6em]
\Rightarrow ~~ \mathrm{num}\big(\mathbb{P}^1_{[C]}\big) + 2 \,\mathrm{num}\big(\mathbb{P}^1_{2[C]} \big) &= \tfrac{1}{2} \, c_2(X) \cdot \big( \mathcal{H}_{(i)} - H_{(i)} \big) 
\\
&= \tfrac{1}{2} \, c_2(X) \cdot \Big( -2 \,,\, \vec{q}_1 + 2\sum_{k=2}^n \vec{q}_k \Big)\,.
\end{aligned}
\label{eq:expl_flops2_other_p1sum}
\ee
\end{proof}

\begin{ex}
Consider a generic CICY threefold in the family
\be
X_{7758} \sim
\left[
\begin{array}{c | c c c }
\mathbb{P}^2 & 2 & 1 & 0 \\
\mathbb{P}^4 & 1 & 2 & 2 \\
\end{array}
\right]
\,.
\ee
One can check that the Euler characteristic and second Chern class of this Calabi-Yau are $\ec(X) = -100$ and $c_2(X) = (36,52)$. The divisor class of $\mathcal{H}_{(i)}$ can be read off from the configuration matrix as $(-1,5)$. The smoothed contracted manifold $\defm{X}$ is
\be
\defm{X} \sim
\begin{array}{c c c c | c c }
\xi  & y_1	& \ldots & y_4 	& \defm{P}_1  & \defm{P}_2 \\
\hline \rule{0pt}{2.5ex}
3  & 1 & \ldots & 1			& 6  & 2  \\
\end{array} \,,
\ee
and one can check that its Euler characteristic is $\ec(\defm{X}) = -256$. With this information we find that
\be
\mathrm{num}\big(\mathbb{P}^1_{[C]}\big) = 62 \,, \quad \mathrm{num}\big(\mathbb{P}^1_{2[C]}\big) = 16 \,.
\ee
\end{ex}
\medskip

\begin{rmk}
One can also compute $\mathrm{num}\big(\mathbb{P}^1_{2[C]}\big)$ directly by using the Giambelli-Thom-Porteous formula, analogously to the computation in \sref{sec:expl_flops_count}. Defining the line bundle sum $F = \bigoplus_{j=2}^{n} \mathcal{O}_{\vec{\mathbb{P}}}(\vec{q}_j)$ on $\vec{\mathbb{P}}[y]$, and writing $\mathcal{F}$ for the restriction of $F$ to $Y \sim \big[\begin{array}{c | c c c} \vec{\mathbb{P}}[y] &  \vec{q}_{n+1} & \hdots & \vec{q}_K \end{array}\big]$, the number of $\mathbb{P}^1$s with class $2C$ is
\be
\mathrm{num}\big(\mathbb{P}^1_{2[C]}\big) = 
\left|
\begin{array}{c c c}
c_1(\mathcal{F}) & c_2(\mathcal{F}) & c_3(\mathcal{F}) \\
c_0(\mathcal{F}) & c_1(\mathcal{F}) & c_2(\mathcal{F}) \\
0 & c_0(\mathcal{F}) & c_1(\mathcal{F}) 
\end{array}
\right| 
=
\left|
\begin{array}{c c c}
c_1(F) & c_2(F) & c_3(F) \\
c_0(F) & c_1(F) & c_2(F) \\
0 & c_0(F) & c_1(F) 
\end{array}
\right| 
\, \cdot
\prod_{j=n+1}^{K} \Big( \, \vec{q}_j \cdot \vec{H} \, \Big) \,.
\ee
In the last expression all intersections are taken on $\vec{\mathbb{P}}[y]$, and the product factor is a series of intersections which implement the restriction to $Y \sim \big[\begin{array}{c | c c c} \vec{\mathbb{P}}[y] &  \vec{q}_{n+1} & \hdots & \vec{q}_K \end{array}\big]$, in which $\vec{H}$ is a list of the hyperplane classes of each of the projective spaces in $\vec{\mathbb{P}}[y]$.  
\end{rmk}

\begin{ex}Consider a generic CICY threefold in the family
\be
X_{7759} \sim
\left[
\begin{array}{c | c c c c }
\mathbb{P}^3 & 2 & 1 & 1 & 0\\
\mathbb{P}^4 & 1 & 1 & 1 & 2\\
\end{array}
\right]
\,,
\ee
so that $F = \mathcal{O}_{\mathbb{P}^4}(1) \oplus \mathcal{O}_{\mathbb{P}^4}(1)$ and $\prod_{j=n+2}^{K} \big(\vec{q}_j \cdot \vec{H}\big) = 2H$. The Chern classes of $F$ are then
\be
c_0(F) = 1 \,, \quad c_1(F) = 2H \,, \quad c_2(F) = H^2 \,, \quad c_3(F) = 0 \,.
\ee
and the number of collapsing curves in class $2[C]$ is 
\be
\mathrm{num}(\mathbb{P}^1_{2[C]}) =
\left|
\begin{array}{c c c}
2H & H^2 & 0 \\
1 & 2H & H^2 \\
0 & 1 & 2H 
\end{array}
\right|  \cdot (2H) = 8 \,.
\ee
One can also check using \eref{eq:expl_flops2_euldif} or \eref{eq:expl_flops2_other_p1sum} that $\mathrm{num}(\mathbb{P}^1_{[C]}) = 70$. Lastly we note that for this Calabi-Yau threefold, both rows are Type~2, so both boundaries of the K\"ahler cone correspond to flops to isomorphic manifolds, and the Calabi-Yau admits two distinct descriptions as a small resolution of a singular double cover. Focusing on the flop on the second row, one finds instead $\mathrm{num}(\mathbb{P}^1_{[C]}) = 72$ and $\mathrm{num}(\mathbb{P}^1_{2[C]}) = 26$.
\end{ex}

\providecommand{\href}[2]{#2}\begingroup\raggedright\endgroup


\begin{thebibliography}{10}

\bibitem{Kawamata:1988}
Y.~Kawamata, ``{Crepant blowing-up of 3-dimensional canonical singularities and
  its application to degenerations of surfaces},'' {\em Ann. of Math. (2)}
  {\bfseries 127} (1988) 93--163.

\bibitem{Atiyah:1958}
M.~F. Atiyah, ``{On analytic surfaces with double points},'' {\em Proc. Roy.
  Soc. London. Ser. A} {\bfseries 247} (1958) 237--244.

\bibitem{Kollar:1989}
J.~Koll\'ar, ``{Flops},'' {\em Nagoya Math. J.} {\bfseries 113} (1989) 14--36.

\bibitem{Laufer:1981}
H.~B. Laufer, ``{On $\mathbb{CP}^1$ as exceptional set, Recent Developments in
  Several Complex Variables (J. E. Fornaess, ed.)},'' {\em Ann. of Math. Stud}
  {\bfseries 100} (1981) 261--275.

\bibitem{KatzMorrison:1992}
S.~Katz and D.~R. Morrison, ``{Gorenstein threefold singularities with small
  resolutions via invariant theory for Weyl groups},'' {\em J. Algebraic Geom.}
  {\bfseries 1} (1992) 449--530, \href{http://arxiv.org/abs/9202002}{{\ttfamily
  arXiv:9202002 [alg-geom]}}.

\bibitem{wall}
C.~T.~C. Wall, ``Classification problems in Differential Topology V. On certain 6-manifolds," {\em Invent. math.} {\bfseries 1} (1966) 355–-374.

\bibitem{Candelas:1987kf}
P.~Candelas, A.~Dale, C.~Lutken, and R.~Schimmrigk, ``{Complete Intersection
  Calabi-Yau Manifolds},''
  \href{http://dx.doi.org/10.1016/0550-3213(88)90352-5}{{\em Nucl. Phys. B}
  {\bfseries 298} (1988) 493}.

\bibitem{Green:1988wa}
P.~S. Green and T.~Hubsch, ``{Possible Phase Transitions among Calabi-Yau
  Compactifications},''
  \href{http://dx.doi.org/10.1103/PhysRevLett.61.1163}{{\em Phys. Rev. Lett.}
  {\bfseries 61} (1988) 1163}.

\bibitem{Green:1988bp}
P.~S. Green and T.~Hubsch, ``{Connecting Moduli Spaces of Calabi-yau
  Threefolds},'' \href{http://dx.doi.org/10.1007/BF01218081}{{\em Commun. Math.
  Phys.} {\bfseries 119} (1988) 431--441}.

\bibitem{Green:1987cr}
P.~S. Green, T.~Hubsch, and C.~A. Lutken, ``{All Hodge Numbers of All Complete
  Intersection Calabi-Yau Manifolds},''
  \href{http://dx.doi.org/10.1088/0264-9381/6/2/006}{{\em Class. Quant. Grav.}
  {\bfseries 6} (1989) 105--124}.

\bibitem{Anderson:2017aux}
L.~B. Anderson, X.~Gao, J.~Gray, and S.-J. Lee, ``{Fibrations in CICY
  Threefolds},'' \href{http://dx.doi.org/10.1007/JHEP10(2017)077}{{\em JHEP}
  {\bfseries 10} (2017) 077}, \href{http://arxiv.org/abs/1708.07907}{{\ttfamily
  arXiv:1708.07907 [hep-th]}}.

\bibitem{Brodie:2020fiq}
C.~R. Brodie, A.~Constantin, and A.~Lukas, ``{Flops, Gromov-Witten invariants
  and symmetries of line bundle cohomology on Calabi-Yau three-folds},''
  \href{http://dx.doi.org/10.1016/j.geomphys.2021.104398}{{\em J. Geom. Phys.}
  {\bfseries 171} (2022) 104398},
  \href{http://arxiv.org/abs/2010.06597}{{\ttfamily arXiv:2010.06597
  [hep-th]}}.

\bibitem{Brodie:2021ain}
C.~R. Brodie, A.~Constantin, A.~Lukas, and F.~Ruehle, ``{Swampland conjectures
  and infinite flop chains},''
  \href{http://dx.doi.org/10.1103/PhysRevD.104.046008}{{\em Phys. Rev. D}
  {\bfseries 104} no.~4, (2021) 046008},
  \href{http://arxiv.org/abs/2104.03325}{{\ttfamily arXiv:2104.03325
  [hep-th]}}.
  
\bibitem{Brodie:2021nit}
C.~R. Brodie, A.~Constantin, A.~Lukas, and F.~Ruehle, ``{Geodesics in the
  extended K\"ahler cone of Calabi-Yau threefolds},''
  \href{http://dx.doi.org/10.1007/JHEP03(2022)024}{{\em JHEP}
  {\bfseries 3} (2022) 024},
  \href{http://arxiv.org/abs/2108.10323}{{\ttfamily arXiv:2108.10323
  [hep-th]}}


\bibitem{Brodie:2021zqq}
C.~Brodie, A.~Constantin, J.~Gray, A.~Lukas and F.~Ruehle,
``{Recent Developments in Line Bundle Cohomology and Applications to String Phenomenology},''  \href{http://arxiv.org/abs/2112.12107}{{\ttfamily arXiv:2112.12107 [hep-th]}}.

\bibitem{Constantin:2022jyd}
A.~Constantin,
``{Intelligent Explorations of the String Theory Landscape},'' \href{http://arxiv.org/abs/2204.08073}{{\ttfamily arXiv:2204.08073 [hep-th]}}.


\bibitem{Wilson1997FlopsTI}
P.~M.~H.~Wilson, ``{Flops, Type III contractions and Gromov-Witten invariants 
  on Calabi-Yau threefolds},''
  In : New Trends in Algebraic Geometry (ed.\ K.~Hulek, F.~Catanese, C.~Peters \&
  M.~Reid), pp. 465-484. CUP, Cambridge, 1999.

\bibitem{KollarMori:1998}
J.~Koll\'ar and S.~Mori, {\em {Birational geometry of algebraic varieties}}.
\newblock CUP, Cambridge, 1998.

\bibitem{katz1996}
S.~Katz, D.~R. Morrison, and M.~Plesser, ``Enhanced gauge symmetry in type 11
  string theory,'' \href{http://dx.doi.org/10.1016/0550-3213(96)00331-8}{{\em
  Nuclear Physics B} {\bfseries 477} no.~1, (Oct, 1996) 105–140}.

\bibitem{Jockers:2012zr}
H.~Jockers, V.~Kumar, J.~M.~Lapan, D.~R.~Morrison, and M.~Romo,
``{Nonabelian 2D Gauge Theories for Determinantal Calabi-Yau Varieties}'',
  \href{http://dx.doi.org/10.1007/JHEP11(2012)166}{{\em JHEP}
  {\bfseries 3} (2012) 166},
  \href{http://arxiv.org/abs/1205.3192}{{\ttfamily arXiv:1205.3192
  [hep-th]}}.
  
\bibitem{Candelas:1989js}
P.~Candelas and X.~C. de~la Ossa, ``{Comments on Conifolds}",
  \href{http://dx.doi.org/10.1016/0550-3213(90)90577-Z}{{\em Nucl. Phys. B}
  {\bfseries 342} (1990) 246--268}.

\bibitem{Ottem2013BirationalGO}
J.~C. Ottem, ``Birational geometry of hypersurfaces in products of projective
  spaces,'' {\em Mathematische Zeitschrift} {\bfseries 280} (2013) 135--148.

\end{thebibliography}
\end{document}